\documentclass[useAMS,usenatbib]{mnras}
\usepackage{graphicx, bm, amssymb, xcolor}
\usepackage{verbatim}
\usepackage[all]{hypcap}
\usepackage{xspace}
\usepackage{multirow,bigdelim}
\usepackage[fleqn]{amsmath}
\usepackage{threeparttable}
\usepackage{epsfig}
\usepackage{aas_macros}
\usepackage{natbib}
\usepackage{times,txfonts}
\usepackage{morefloats}
\usepackage{tabularx}
\usepackage{lscape}
\usepackage{hyperref}
\usepackage{xcolor}
\usepackage{orcidlink}
\usepackage{adjustbox}
\newcommand{\mic}{\,$\mu$m} 
\newcommand{\Ha}{H$\alpha$\xspace}

\newcommand{\HI}{H\,{\sc i}\xspace}     
\newcommand{\HII}{H\,{\sc ii}\xspace}     
\newcommand{\NII}{{\rm [N}$\,${\sc ii}{\rm ]}\xspace}
\newcommand{\coone}{\mbox{\rm CO($1\text{--}0$)}\xspace} 
\newcommand{\cotwo}{\mbox{\rm CO($2\text{--}1$)}\xspace} 
\newcommand{\tHa}{$t_{\rm H\alpha}$}   
\newcommand{\tir}{$t_{\rm 24\,\mu m}$}
\newcommand{\tCO}{$t_{\rm CO}$}   
\newcommand{\Msun}{\mbox{M$_\odot$}}

\newcommand{\pyr}{\mbox{${\rm yr^{-1}}$}}

\newcommand{\pc}{\mbox{${\rm~pc}$}}

\newcommand{\kpc}{\mbox{${\rm kpc}$}}
\newcommand{\kms}{\mbox{${\rm km}~{\rm s}^{-1}$}}

\newcommand{\vfb}{$v_{\rm fb}$}
\newcommand{\esf}{$\epsilon_{\rm sf}$}

\newcommand{\Heisenberg}{{H\scriptsize{EISENBERG}\xspace}}
\newcommand{\be}{\begin{equation}}
\newcommand{\ee}{\end{equation}}
\newcommand{\bea}{\begin{eqnarray}}
\newcommand{\eea}{\end{eqnarray}}

\newcommand{\appropto}{\mathrel{\vcenter{
  \offinterlineskip\halign{\hfil$##$\cr
    \propto\cr\noalign{\kern1pt}\sim\cr\noalign{\kern-2pt}}}}}

\usepackage{etoolbox}

\title[]{\vspace{-6mm}On the duration of the embedded phase of star formation \vspace{-5mm}}

\author{Jaeyeon~Kim\orcidlink{0000-0002-0432-6847},$^1$\thanks{E-mail: \href{kim@uni-heidelberg.de}{kim@uni-heidelberg.de}}
M\'{e}lanie~Chevance\orcidlink{0000-0002-5635-5180},$^1$
J.~M.~Diederik~Kruijssen\orcidlink{0000-0002-8804-0212},$^1$
\newauthor
Andreas~Schruba,$^2$ 
Karin Sandstrom,$^3$
Ashley~T.~Barnes\orcidlink{0000-0003-0410-4504},$^{4}$ 
Frank Bigiel,$^4$
\newauthor
 Guillermo A.~Blanc,$^{5,6}$
Yixian Cao,$^7$ 
Daniel~A.~Dale,$^8$ 
Christopher M.~Faesi,$^{9}$ 
\newauthor
Simon~C.~O.~Glover,$^{10}$
Kathryn~Grasha\orcidlink{0000-0002-3247-5321},$^{11}$
Brent~Groves,$^{12}$ 
Cinthya Herrera, $^{13}$
\newauthor
Ralf~S.~Klessen\orcidlink{0000-0002-0560-3172},$^{10,14}$
Kathryn~Kreckel,$^{1}$
Janice~ C.~Lee,$^{15}$
Adam~K.~Leroy,$^{16}$ 
\newauthor
J\'{e}r\^{o}me Pety\orcidlink{0000-0003-3061-6546},$^{13,17}$ 
Miguel~Querejeta,$^{18}$ 
Eva Schinnerer,$^{19}$
Jiayi~Sun\orcidlink{0000-0003-0378-4667},$^{16}$
\newauthor
Antonio Usero,$^{18}$
Jacob L. Ward,$^{1}$
Thomas G. Williams\orcidlink{0000-0002-0012-2142}$^{19}$      
\\\\
Affiliations are listed at the end of the paper
\vspace{-4mm}}
\begin{document}

\date{Accepted X{\sevensize xxxx} XX. Received X{\sevensize xxxx} XX; in original form 2020 November 30\vspace{-3mm}}

\pagerange{\pageref{firstpage}--\pageref{lastpage}} \pubyear{2020}

\maketitle

\label{firstpage}

\begin{abstract}
Feedback from massive stars plays a key role in molecular cloud evolution. After the onset of star formation, the young stellar population is exposed by photoionization, winds, supernovae, and radiation pressure from massive stars. Recent observations of nearby galaxies have provided the evolutionary timeline between molecular clouds and exposed young stars, but the duration of the embedded phase of massive star formation is still ill-constrained. We measure how long massive stellar populations remain embedded within their natal cloud, by applying a statistical method to six nearby galaxies at $20{-}100~\pc$ resolution, using CO, \textit{Spitzer} 24$\rm\,\mu m$, and H$\alpha$ emission as tracers of molecular clouds, embedded star formation, and exposed star formation, respectively. We find that the embedded phase (with CO and 24$\rm\,\mu m$~emission) lasts for $2{-}7$~Myr and constitutes $17{-}47\%$ of the cloud lifetime. During approximately the first half of this phase, the region is invisible in H$\alpha$, making it heavily obscured. For the second half of this phase, the region also emits in H$\alpha$ and is partially exposed. Once the cloud has been dispersed by feedback, 24$\rm\,\mu m$ emission no longer traces ongoing star formation, but remains detectable for another $2{-}9$~Myr through the emission from ambient CO-dark gas, tracing star formation that recently ended. The short duration of massive star formation suggests that pre-supernova feedback (photoionization and winds) is important in disrupting molecular clouds. The measured timescales do not show significant correlations with environmental properties (e.g.\ metallicity). Future JWST observations will enable these measurements routinely across the nearby galaxy population.

\end{abstract}


\section{Introduction} \label{sec:intro}
Massive stars ($>8~{\rm M}_{\odot}$) form in the densest regions of molecular clouds. Once formed, these stars emit large quantities of ionizing photons creating \HII regions and generate strong winds, which together alter the structure of their birth clouds and the surrounding interstellar medium. After a relatively short lifetime ($4-20$~Myr; \citealp{leitherer14, barnes20, chevance20_fb}), these massive stars  die in supernova explosions injecting energy and momentum into their surroundings. Theoretical studies of giant molecular clouds (GMCs) indicate that these feedback processes are responsible for freeing the young stars from their parental clouds and destroying the GMCs \citep[see e.g.][for recent reviews]{krumholz14,dale15, chevance20_rev}. However, it is still debated which feedback mechanisms efficiently disrupt the birth clouds and which affect the diffuse interstellar medium on large scales (e.g. \citealp{lucas20}; A.~Barnes et al. in preparation).

Observationally, several studies have constrained the timescale for GMC destruction by stellar feedback in the Milky Way and in nearby galaxies using optical and UV star formation tracers, which are sensitive to recent, not heavily obscured star formation. Rapid dispersion of GMCs, within a cloud dynamical timescale ($\leq$10~Myr), has been suggested based on the age distributions of stars in nearby star-forming regions and young stellar associations \citep{elmegreen00, hartmann01}. For GMCs in the Milky Way, M33, and the Large Magellanic Cloud (LMC), somewhat longer feedback timescales of $10-20$~Myr have been proposed by classifying molecular clouds into different types based on the \mbox{(non-)}\linebreak[0]{}existence of their star formation activity \citep{engargiola03, blitz07, kawamura09, miura12, corbelli17}. Such methods have limitations because individual GMCs and star-forming regions need to be resolved. In addition, most of these studies only constrain the duration of the cloud dispersal after the young stars have become partially exposed. The exact role of each different feedback mechanism remains ambiguous, because the total duration of the embedded phase of star formation, including a heavily obscured phase, has not been quantified. 

Studies of star-forming regions in the Milky Way and some nearby galaxies show that the embedded phase of massive star formation lasts for $2-5$ Myr, where the duration of the heavily obscured phase is found to be $\sim 0.1 - 2$ Myr \citep{lada03,  whitmore14, corbelli17}. During the earliest stage of star formation, young massive stars are still embedded in their natal gas. As a result, H$\alpha$ emission is heavily or partially obscured due to the extinction provided by the dust in dense gas surrounding the young stars. Despite this, on-going star formation is detectable using mid-infrared, hydrogen infrared recombination lines, radio recombination lines, and free-free radio continuum emission \citep{lockman89,kennicutt98_rev, calzetti05, kennicutt07, prescott07, murphy11,  vutisalchavakul13, querejeta19}. In particular, the 24$\rm\,\mu m$ emission in the mid-infrared observed by the Multiband Imaging Photometer (MIPS) aboard the \textit{Spitzer} Space Telescope provides an unbiased tracer of embedded massive star formation (e.g. see \citealp{kennicutt12} for review), as long as the star-forming region has a fully sampled initial mass function \citep{vutisalchavakul13}. 24$\rm\,\mu m$ emission originates from stochastically heated small dust grains that do not require ionizing photons to be excited but do empirically correlate with tracers of massive star formation and so can be used as a tracer of the presence of large amounts of dust-reprocessed photospheric light from massive stars. Therefore, the 24$\rm\,\mu m$ emission is assumed to turn on only once massive stars are present. However, the radiation field from older stars (i.e. late-type B~stars with an age of $\lesssim$100 Myr) can make a non-negligible contribution to the dust heating and thus the mid-infrared dust emission \citep{draine07, verley09, leroy12}. The exact contribution at 24$\rm\,\mu m$ wavelength is found to vary strongly between galaxies (e.g. 85\% in M31 by \citealp{viaene17}, 20\% in M33 by \citealp{williams19}, and up to around $>60\%$ within the Galactic centre of the Milky Way, see \citealp{koepferl15}).

Following the theoretical model of star formation by \citet{schmidt59}, it is now observationally well-known that on galactic scales there is a tight correlation between the molecular gas surface density and the SFR surface density (the ``star formation relation''; \citealp{silk97, Kennicutt98, bigiel08, leroy13}). However, this relation is also observed to break down on scales smaller than $\sim$1~kpc \citep[e.g.][]{onodera10,schruba10,ford13,leroy13,kreckel18,williams18,kruijssen19,schinnerer19}. This breakdown is caused by the small-scale de-correlation between GMCs and young stellar regions on sub-kpc scales and can be explained by assuming that individual regions in a galaxy follow independent lifecycles, during which clouds assemble, form stars and get disrupted by feedback \citep{schruba10, feldmann11, kruijssen14}. 

\citet{kruijssen14} and \citet{kruijssen18} developed a formalism that translates the observed de-correlation quantitatively into the evolutionary timeline of GMCs from cloud formation to subsequent star formation, and finally cloud dispersal. In brief, this method uses the spatial variation of the gas-to-SFR flux ratio observed at different spatial scales ranging from cloud ($\sim$100~pc) to galactic ($\sim$1 kpc) scales  for apertures placed on either gas peaks or SFR tracer peaks. These measurements are then used to determine the duration of each phase of the evolutionary cycle: the cloud lifetime, the timescale a SFR tracer is visible, and the phase during which both molecular gas and SFR tracers overlap, which represents the duration of massive star formation as well as the timescale for molecular gas to be removed or dissociated by feedback. This method does not rely on high angular resolution as much as previous methods, using stellar ages or relative fractions of GMCs with and without internal star formation activity, since it only requires the mean separation length between regions undergoing independent evolution ($100-200$~pc; \citealp{chevance20}) to be resolved instead of resolving individual star-forming regions ($10-50$~pc; \citealp{kawamura09}).

This method has been used to characterise the evolutionary timelines between molecular gas and exposed young stellar populations using CO and H$\rm \alpha$ observations. The first applications of this method to a number of galaxies covering a large range of galactic environments (\citealt{kruijssen19,chevance20_fb, chevance20, hygatePhD, ward20, zabel20}) have shown that GMCs live for $10-30$~Myr. CO and H$\alpha$ emission are found to be coincident for $1-5$~Myr, during which time feedback from the young stellar population disperses the molecular gas of their birth clouds. Considering that supernova explosions are expected $4-20$~Myr after massive stars are formed \citep{leitherer14, chevance20_fb}, the short duration of overlapping CO and H$\alpha$ emission suggests that pre-supernova feedback, such as photoionization and stellar winds, is important for disrupting star-forming molecular clouds \citep[also see \citealt{barnes20}]{chevance20_fb}. \citet{ward20_hI} have also used this method and extended the evolutionary timeline of star-forming regions by incorporating \HI emission to trace atomic gas. The measured atomic gas cloud lifetime in the LMC is $\sim$50~Myr and almost no overlap with the exposed star-forming phase is detected. 

In this paper, we go a step further in the characterisation of the GMC evolutionary lifecycle and use CO, 24$\rm\,\mu m$, and H$\alpha$ emission as tracers for molecular gas, embedded star formation, and exposed star formation, respectively, for six nearby galaxies (IC\,342, LMC, M31, M33, M51, and NGC\,300). Previous applications of the statistical method to the same galaxies have focused on characterising the GMC lifecycle using CO emission as a tracer of the molecular gas and H$\alpha$ emission as a tracer of the young massive stars (NGC\,300: \citealt{kruijssen19}; M51: \citealt{chevance20}; M33: \citealt{hygatePhD}; and the LMC: \citealt{ward20}). We derive novel measurements of the GMC lifetimes in IC\,342 and M31 in this paper (see Appendix~\ref{sec:app_tco}); these are based on CO and H$\alpha$ observations presented in \citet{schruba21_ic342,schruba21_m31}. Employing \textit{Spitzer} 24$\rm\,\mu m$ observations allows us to probe the earliest phase of star formation where the stars are still heavily obscured, and quantify how long it takes young star-forming regions to emerge from their natal cloud (at which point the \HII regions created by their ionizing radiation become visible).

The structure of this paper is as follows. In Section~\ref{sec:data} and Appendix~\ref{sec:app_map}, we describe the observational data used in our analysis. In Section~\ref{sec:method}, we summarise the statistical method used here and describe the associated input parameters for the galaxies in our sample. This is complemented by Appendix~\ref{sec:app_tco}, where we present in more detail the application of this method to IC\,342 and M31. In Section~\ref{sec:result}, we then present the derived duration of the embedded massive star-forming phase, which can be separated into a heavily obscured phase of star formation and a partially exposed phase of star formation based on the existence of H$\alpha$ emission. The total duration of the 24$\rm\,\mu m$ emitting phase is also presented. In addition, we explore how the durations of these phases vary with environmental properties, across the small galaxy sample for which these measurements are possible. We discuss the robustness of our results and compare them with the literature in Section~\ref{sec:discussion}. Last, we present our conclusions in Section~\ref{sec:conclusion}.

\begin{table*}
\begin{center}
\caption{Physical and observational properties of our galaxy sample.} \label{tab:obs}
\begin{threeparttable}
\begin{adjustbox}{width=\textwidth}
\begin{tabular}{lccccccccc}
\hline
Galaxy & Stellar mass$^{a}$ & Metallicity$^{b, c}$ & Distance & Inclination & Position angle& CO  & CO  & 24$\rm\,\mu m$ & Spatial \\
&&&&&&observations&resolution &resolution &resolution$^{d}$ \\
&[$\rm log_{10}\,M_{\odot}$]&[$\rm Z/Z_{\odot}$]& [Mpc] & [deg] &[deg]&& [\arcsec]\ & [\arcsec]\ & [pc]\\
\hline
IC\,342& $10.2\pm 0.1$ &$0.90 \pm 0.20$ & 3.45 & 31.0& 42.0& NOEMA  & 3.6 &6.4&107\\
&&&&&&+ IRAM \mbox{30-m}&&&\\
LMC &$9.3\pm 0.1$& $\rm 0.48 \pm 0.03$ & 0.05 & 22.0 &168.0& ATNF & 45 & 6.4 & 11\\
M31 (NGC\,224) &$11.0\pm0.1$& $\rm 0.76 \pm 0.20$ & 0.78 & 77.7 &37.7& CARMA & 5.5&6.4&24 \\
&&&&&&+ IRAM \mbox{30-m}&&&\\
M33 (NGC\,598) &$9.4\pm0.1$& $\rm0.50 \pm 0.06$ & 0.84 & 55.0 &201.1& IRAM \mbox{30-m} & 12&6.4& 49 \\
M51 (NGC\,5194) &$10.7\pm 0.1$&$\rm 1.37 \pm 0.20$ & 8.6 & 21.0&173.0 & PdBI  & 1.1&2.4 & 100\\
&&&&&&+ IRAM \mbox{30-m}&&&\\
NGC\,300&$9.3\pm 0.1$& $\rm0.48 \pm 0.06$ & 2.0 & 42.0 & 111.0 & ALMA & 2.1 &6.4& 62\\
\hline
\end{tabular}
\end{adjustbox}
\begin{tablenotes}
\item[\textit{a}]Adopted from \citet{skibba12} for the LMC and \citet{sick15} for M31, while others are from \citet{leroy19}.
\item[\textit{b}]CO luminosity weighted metallicity over the considered field of view.
\item[\textit{c}]Obtained using $\rm Z/Z_{\odot} =(O/H)/(O/H)_{\odot}$, with the solar oxygen abundance $\rm 12 +log(O/H)_{\odot}=8.69$ \citep{asplund09}.
\item[\textit{d}]Coarsest spatial resolution of the CO and 24$\rm\,\mu m$ maps.
\end{tablenotes}
\end{threeparttable}
\end{center}
\end{table*}

\begin{figure*}
\includegraphics[height=21.6cm]{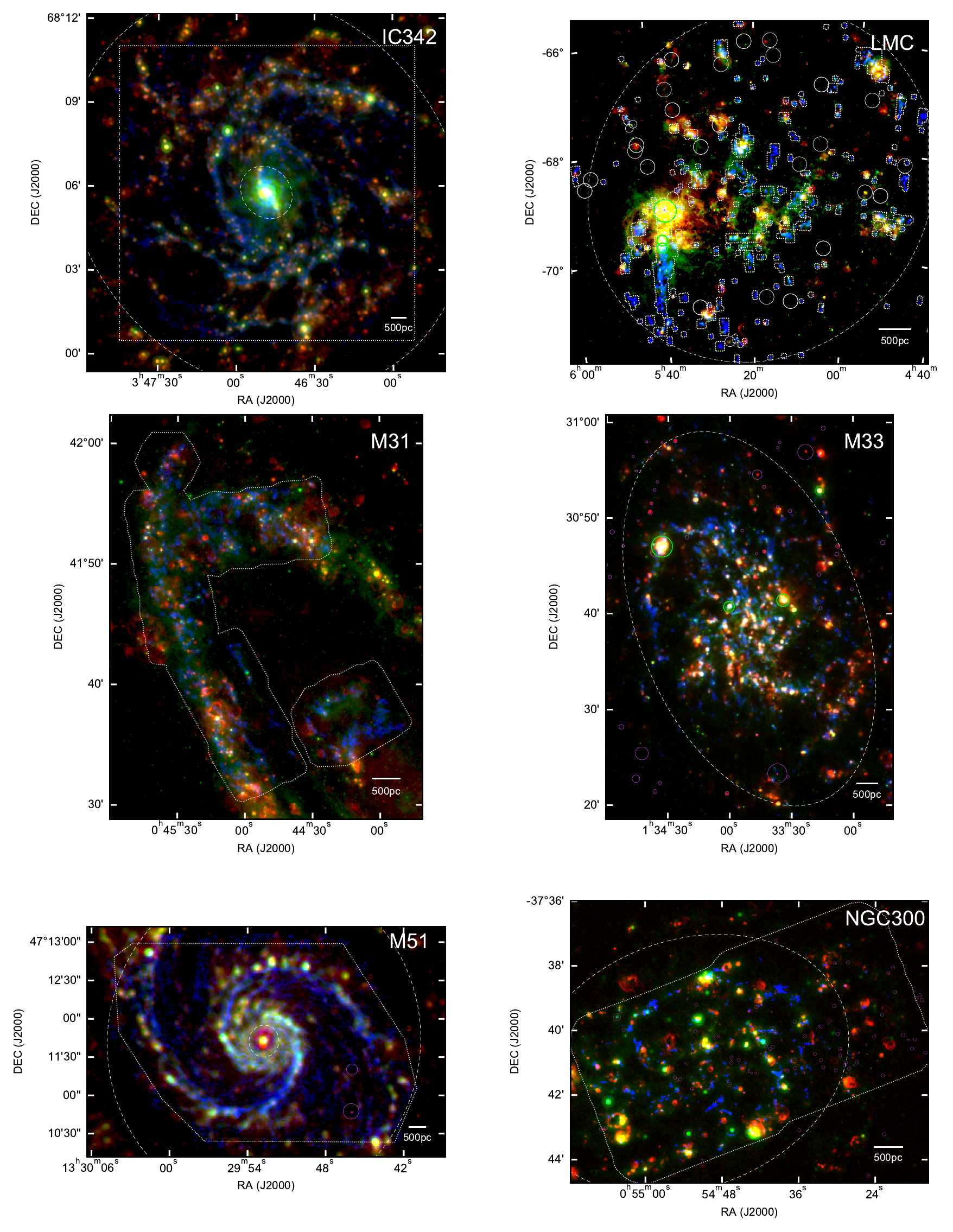}
\caption{Composite three-color images of the six galaxies in our sample. The CO(1-0) emission (CO(2-1) for M33) is presented in blue, {\it Spitzer} MIPS 24$\rm\,\mu m$ in green, and H$\alpha$ in red. Galaxies are sorted by alphabetical order. The range of galactic radii included in the analysis is indicated by the white dashed ellipses. The inner ellipses of IC\,342 and M51 show galaxy centres excluded from our analysis. The white dotted line shows the area where CO emission was observed in each galaxy (excluding the high noise edges of the map) and is not shown for M33 because it is outside the field of view. The massive star-forming regions of 30 Doradus in the LMC and NGC\,604 in M33, as well as bright 24$\rm\,\mu m$ peaks are masked (green circles; see Section~\ref{subsec:mask}). In the LMC, regions that were not targeted by the MAGMA survey but are known to host gas clouds are masked (white solid circles). Foreground stars that were not removed in the image reduction process and image artifacts are also masked (purple circles). A physical scale of 500~pc is shown in each image.} \label{fig:map_rgb}
\end{figure*}

\section{Observational Data}\label{sec:data}
We apply our analysis to six nearby galaxies: IC\,342, the LMC, M31, M33, M51, and NGC\,300. The size of our sample is mostly limited by the angular resolution of the \textit{Spitzer} MIPS observations at 24$\rm\,\mu m$ ($\sim$6.4\arcsec, \citealp{rieke04}; corresponding to $2-110$~pc for the galaxies in our sample, including M51 for which we use a map at higher angular resolution, see Section~\ref{subsec:datasets}) and the fact that we need to resolve at least the separation length between star-forming regions undergoing independent evolutionary lifecycles ($\sim 100-200$~pc) to apply our analysis (see Section~\ref{subsec:robust_kl18} and \citealp{kruijssen18}). Here, we provide a summary of the observational data used to trace the molecular gas (emission from low-\textit{J} CO transitions), embedded massive stars (24$\rm\,\mu m$ emission), and exposed young stellar regions (H$\alpha$ emission). Composite three-color images of the CO, 24$\rm\,\mu m$, and H$\alpha$ maps are presented in Figure~\ref{fig:map_rgb}, whereas the individual CO and 24$\rm\,\mu m$ maps are shown in Appendix~\ref{sec:app_map}. The physical and observational properties of the galaxies in our sample are listed in Table~\ref{tab:obs}.

\subsection{Descriptions of multi-wavelength data sets}\label{subsec:datasets}

\textit{IC\,342.}  We adopt a distance of 3.45~Mpc \citep{wu14}, an inclination of $31.0^{\circ}$, and a position angle of $42.0^{\circ}$ \citep{meidt09}. The adopted metallicity at the galactic centre is $\rm 12+log(O/H)_{0}=8.83 \pm 0.04$, with a radial gradient of $-0.0500 \pm 0.0085~{\rm dex~kpc}^{-1}$, as measured by \citet{pilyugin14} using the strong-line ``counterpart'' method by \citet{pilyugin12}. We use a combination of NOEMA and IRAM \mbox{30-m} observations of the $^{12}$CO ($J{=}1{-}0$) transition (denoted as \coone in the following) from \citet{schruba21_ic342} to trace molecular gas in IC\,342, covering out to 7.7~kpc in galactic radius. The final data cube has a circular beam size of 3.6\arcsec,\ a spectral resolution of 5~\kms,\ and a sensitivity of 135~mK per 5~\kms\ velocity channel. Integrating over 10~\kms,\ this translates to a 5$\sigma$ point source sensitivity of 5$\sigma\rm(M_{H_{2}})\approx 6.1\times10^4~M_{\odot}$ when the \coone-to-$\rm H_{2}$ conversion factor ($\rm\alpha_{CO}$) listed in Table~\ref{tab:input} is assumed. In order to trace embedded star formation, we retrieved \textit{Spitzer} MIPS 24$\rm\,\mu m$ image from the Local Volume Legacy (LVL) Survey \citep{dale09}. In addition to the procedure described in \citet{dale09}, we mask foreground stars with a G-band magnitude $\leq22$~mag using the Gaia DR2 catalog \citep{gaiadr2}, as well as background galaxies via visual inspection of extended and non-circular sources. We then subtract background emission by fitting a plane determined at $2-2.5 R_{25}$ from the galactic centre where $R_{25}$ is the optical radius of the galaxy, obtained from the Lyon Extragalactic Database \citep{paturel03a,paturel03b, makarov14}.  H$\alpha$ emission has been observed with the Mosaic Imager on the Mayall \mbox{4-m} telescope. We utilise calibrated narrow-band H$\alpha$ and R-band images kindly provided by Kimberly Herrmann (private communication). We post-process these images as described in \citet{schruba21_ic342}. In brief, we subtract continuum emission from the H$\alpha$ image, mask Milky Way stars using the Gaia DR2 catalog, subtract a sky background by fitting a $\rm 1^{st}$ order polynomial plane at galactic radii $>10$~kpc, and correct for Galactic extinction adopting $A(\rm H\alpha) = 1.3$~mag which is derived from $\rm E(B{-}V) = 0.494$~mag \citep{schlafly11} and an extinction curve with $R_{\rm V}=3.1$ \citep{cardelli89}\footnote{As explained in the discussion of each galaxy, we do not adopt the same literature for the reddening ($E(B{-}V)$) or the extinction law when correcting for Galactic extinction. This is to follow what has been done previously for each galaxy, in our first applications of the statistical method using CO and H$\alpha$ emission only \citep{kruijssen19, chevance20,hygatePhD, ward20}. Nevertheless, similarly to the \coone-to-$\rm H_{2}$ and SFR conversion factor (see Section~\ref{subsec:input}), we note that our choice of correction factor does not affect our measurements of timescales and region separation length, but only the derived total SFR and the integrated star formation efficiency.}. For all the H$\rm\alpha$ maps used here, we do not attempt to correct for the internal extinction so that it traces exposed star-forming regions. The resulting H$\alpha$ map has an angular resolution of $0.85\arcsec$. Due to limited coverage of the CO survey and blending of bright sources at the galaxy centre, we restrict our analysis to regions where CO observations have been made within galactic radii $1.0-7.7$~kpc, as shown in Figure~\ref{fig:map_rgb}.

\textit{LMC.} We adopt a distance of 50~kpc \citep{pietrzynski19}, an inclination of 22.0$^{\circ}$, and a position angle of 168.0$^{\circ}$ \citep{kim98}. The adopted metallicity at the centre of the LMC is $\rm12+log(O/H)_{0}=8.35 \pm 0.03$, with a radial gradient of $0.0105 \pm 0.0105\ {\rm dex~kpc}^{-1}$, as measured by \citet{toribio17} using a direct measurement of electron temperature ($T_{\rm e}$) from spectra of \HII\ regions (direct $T_{\rm e}$-based method). We employ the \coone data presented in the third data release of the Magellanic Mopra Assessment \citep[MAGMA;][]{wong11, wong17} to trace molecular gas. MAGMA is a CO mapping survey of the LMC and SMC using the Mopra \mbox{22-m} Telescope at the Australia Telescope National Facility. For the LMC, the observations were conducted as a follow-up study of the NANTEN survey \citep{fukui08} by targeting a subset of previously identified molecular clouds ($\sim$160 out of 272 clouds), with an improved resolution in order to resolve the GMCs ($\sim$11~pc). In Figure~\ref{fig:map_rgb}, white dotted lines show the coverage of the MAGMA survey and solid white circles indicate regions where molecular gas has been detected by the NANTEN survey, but not further targeted with the MAGMA survey. We exclude these white circled regions, where we know GMCs exist, from our analysis\footnote{Using the cloud catalogue
from the NANTEN survey \citep{fukui08}, \citet{ward20} have modelled the CO emission from the GMCs that were not observed by MAGMA and have shown that excluding these regions from our analysis has a negligible effect on our measurements.}. However, we still include other regions not observed by the MAGMA survey, which might have diffuse and faint CO emission. We note that the inclusion of these sight lines, not covered by the MAGMA survey, should have a negligible impact on our results because diffuse and faint emission is eventually removed in our analysis through a filtering process (see Section~\ref{sec:method}). Moreover, the MAGMA survey covers most ($\sim$80\%) of the total CO emission from the LMC observed by the NANTEN survey \citep{wong11}. The observed CO emission from the MAGMA survey also shows good agreement with the molecular gas map produced using dust continuum emission modelling \citep{jameson16}, indicating that we cover most of the emission from GMCs. The resulting angular resolution of the MAGMA CO data is $45\arcsec$, and the sensitivity is $0.3$~K per $0.526$~\kms\ velocity channel. Integrating over 10~\kms,\ this translates to a 5$\sigma$ point source sensitivity of 5$\sigma\rm(M_{H_{2}})\approx 2.2\times10^3~M_{\odot}$ (assuming the $\rm\alpha_{CO}$ from Table~\ref{tab:input}). The \textit{Spitzer} MIPS 24$\rm\,\mu m$ image is from the SAGE project \citep{meixner06}, covering $7^{\circ} \times 7^{\circ}$ of the galaxy. The continuum subtracted H$\alpha$ map is from the Southern \mbox{H-Alpha} Sky Survey Atlas (SHASSA; \citealt{gaustad01}) and has a resolution of $48\arcsec$. We correct for Galactic extinction using $A(\rm H\alpha) = 0.16$~mag, which is derived from $\rm E(B{-}V) = 0.06$~mag \citep{staveleysmith03} and an extinction curve with $R_{\rm V}=3.1$ \citep{cardelli89}. For the analysis here, we include emission from $0-3$~\kpc\ in galactic radius, where the outer boundary is indicated in Figure~\ref{fig:map_rgb}.

\textit{M31.} We adopt a distance of $0.78$~Mpc \citep{dalcanton12}, an inclination of $37.7^{\circ}$, and a position angle of $77.7^{\circ}$ \citep{corbelli10}. The adopted metallicity at the galactic centre is $\rm12+log(O/H)_{0} = 8.8\pm0.1$, with a radial gradient of $-0.022\pm0.014\ {\rm dex~kpc}^{-1}$, as measured by \citet{zurita12} using the strong-line calibration from \citet{pilyugin01}. To trace molecular gas, we use \coone data first appeared in \citet{calduprimo16}, with full details presented in \citet{schruba21_m31}. This data is obtained by combining CARMA interferometry data and IRAM \mbox{30-m} data, the latter from \citet{nieten06}. The CARMA observations cover $87~{\rm kpc}^{2}$ of M31's star-forming disc at galactic radii of $6-13$~kpc. They have an angular resolution of $5.5\arcsec$, a spectral resolution of $2.5$~\kms, and a sensitivity of $175$~mK per $2.5$~\kms\ velocity channel. Integrating over 10~\kms,\ this translates to a 5$\sigma$ point source sensitivity of 5$\sigma\rm(M_{H_{2}})\approx 7.3\times10^3~M_{\odot}$ (assuming the $\rm\alpha_{CO}$ from Table~\ref{tab:input}). We utilise the velocity masked moment-zero map, which is designed to be flux-complete (see \citealt{schruba21_m31} for details). We employ the \textit{Spitzer} MIPS 24$\rm\,\mu m$ map presented in \citet{gordon06}. This map is already background subtracted, and we refer the reader to the original paper for more information on the data reduction procedure. The H$\alpha$ emission map is discussed in \citet{schruba21_m31} and has been created from calibrated narrow-band H$\alpha$ and R-band images from the Local Group Galaxies Survey (LGGS; \citealp{massey06}). The observations were carried out by the Mosaic Imager on the Mayall \mbox{4-m} telescope. The calibrated data were post-processed as described in \citet{schruba21_m31}, which include H$\alpha$ continuum subtraction, masking of Milky Way stars using the Gaia DR2 catalog \citep{gaiadr2}, a sky background subtraction by fitting a plane at galactic radii $>20$~kpc, and a correction for the contamination by \NII by assuming that both \NII lines contribute 35\% of the total H$\rm\alpha$ emission, following \citet{azimlu11}. The Galactic extinction is corrected by adopting a factor $A({\rm H\alpha}) = 0.14$~mag, which is derived from $\rm E(B{-}V) = 0.05$~mag \citep{schlafly11} and an extinction curve with $R_{\rm V}=3.1$ \citep{cardelli89}. The resulting H$\alpha$ map has an angular resolutions of $1.5\arcsec$. We perform our analysis on the field of view spanned by the CO observations.

\textit{M33.} We adopt a distance of $0.84$~Mpc \citep{gieren13}, an inclination of $55.08^{\circ}$, and a position angle of $201.1^{\circ}$ \citep{koch18}. The adopted metallicity at the galactic centre is $\rm 12+log(O/H)_{0} = 8.48\pm0.04$, with a radial gradient of $-0.042\pm0.010\ {\rm dex~kpc}^{-1}$, as measured by \citet{bresolin11} using a direct $T_{\rm e}$-based method. We use the $^{12}$CO($J{=}2{-}1$) transition (denoted as \cotwo in the following) data presented in \citet{gratier10} and \citet{druard14} to trace molecular gas. The observations were carried out using the HEterodyne Receiver Array \citep[HERA;][]{schuster04} on the IRAM \mbox{30-m} telescope covering the galaxy out to radii of 7~\kpc. The resulting angular resolution is 12\arcsec\ and the average noise level is 20 mK per 2.6 \kms\ velocity channel. Integrating over 10~\kms,\ this noise level translates to a 5$\sigma$ point source sensitivity of 5$\sigma\rm(M_{H_{2}})\approx 6.2\times10^3~M_{\odot}$ (assuming the $\rm\alpha_{CO}$ from Table~\ref{tab:input}). We retrieve \textit{Spitzer} MIPS 24$\rm\,\mu m$ image from the Local Volume Legacy (LVL) Survey \citep{dale09}. We then apply the same post-processing procedures as described above for IC\,342. The narrow-band H$\alpha$ data is from \citet{greenawalt98}. The observations were carried out using the Burrell-Schmidt \mbox{0.6-m} telescope at the Kitt Peak National Observatory (KPNO). Detailed information about the image reduction process can be found in \citet{hoopes00}. The Galactic extinction is corrected by using $A({\rm H\alpha}) = 0.1$~mag, obtained from $\rm E(B{-}V)=0.0413$~mag \citep{schlegel98} and an extinction curve with $R_{\rm V}=3.1$ \citep{fitzpatrick07}. The resolution of the H$\alpha$ emission map is $2.0\arcsec$. We restrict our analysis to galactocentric radii $\leq5$~kpc, as outlined in Figure~\ref{fig:map_rgb}. 

\textit{M51.} We adopt a distance of $8.6$~Mpc \citep{bradley09}, an inclination of $21.0^{\circ}$, and a position angle of $173.0^{\circ}$ \citep{colombo14}. The adopted metallicity at the galactic centre is $\rm 12+log(O/H)_{0}=8.88 \pm 0.053$, with a radial gradient of $-0.0223\pm0.0037\ {\rm dex~kpc}^{-1}$, as measured by \citet{pilyugin14} using the strong-line ``counterpart'' method \citep{pilyugin12}. We use the \coone data of the inner $10 \times 6~{\rm kpc}^2$ of the M51 presented in \citet{pety13} as part of the PdBI Arcsecond Whirlpool Survey (PAWS; \citealt{schinnerer13}). The surveyed region is visible in Figure~\ref{fig:map_rgb}. The PdBI observations were carried out using A, B, C, and D configurations. The IRAM \mbox{30-m} telescope was used to recover emission at low spatial frequencies. The final data have an angular resolution of $1.1\arcsec$ and a sensitivity of $0.39$~K per 5~\kms\ velocity channel. Integrating over 10~\kms,\ this noise level translates to a 5$\sigma$ point source sensitivity of 5$\sigma\rm(M_{H_{2}})\approx 8.9\times10^4~M_{\odot}$ (assuming the $\rm\alpha_{CO}$ from Table~\ref{tab:input}). The integrated intensity map was created by applying a mask to the data cube as described in \citet{pety13}. Due to the limited resolution of \textit{Spitzer} MIPS 24$\rm\,\mu m$ imaging ($6.4\arcsec$; \citealp{rieke04}), we can in principle only apply our method to galaxies closer than $\sim$5~Mpc. However, using the higher resolution ($2.4\arcsec$) 24$\rm\,\mu m$ map created by \citet{dumas11}, we are able to expand the application of our method to M51, located at $8.6$~Mpc. This map was created by applying the HiRes deconvolution algorithm \citep{backus05} to the 5th \textit{Spitzer} Infrared Nearby Galaxies Survey (SINGS; \citealt{kennicutt03}) data delivery (see \citealt{dumas11} for more details). However, we note that the artifacts introduced by the deconvolution algorithm could potentially bias our analysis, especially for timescale-related quantities, by modifying the distribution of the 24$\rm\,\mu m$ flux around bright peaks, limiting the interpretation of our results for this galaxy. The H$\alpha$ emission map is also from SINGS \citep{kennicutt03}. The observations were carried out using the KPNO \mbox{2.1-m} telescope with the CFIM imager. The map is corrected for Galactic extinction adopting a correction factor $A({\rm H\alpha}) = 0.08$~mag, obtained from $E(B{-}V)=0.03$~mag \citep{schlafly11} and an extinction curve with $R_{\rm V}=3.1$ \citep{fitzpatrick99}.We also correct for the contamination by \NII lines by scaling the map by a factor of 0.7. The resolution of the H$\alpha$ emission map is $1.83\arcsec$. Because we lack CO observations of the outer galaxy and sources at the galaxy centre are affected by crowding and contamination from active galactic nucleus, we restrict our analysis to the field of view of the CO observations, and within galactic radii of $0.51-5.35$~kpc, as indicated in Figure~\ref{fig:map_rgb}.

\textit{NGC\,300.} We adopt a distance of $2.0$~Mpc \citep{dalcanton09}, an inclination of $42.0^{\circ}$, and a position angle of $111.0^{\circ}$ \citep{westmeier11}. We adopt a metallicity of $\rm 12+log(O/H)_{0} = 8.46\pm0.05$ at the galactic centre and a radial gradient of $-0.056\pm0.015\ {\rm dex~kpc}^{-1}$, as measured by \citet{toribio16} using a direct $T_{\rm e}$-based method. We employ ALMA observations of the \coone transition, from ALMA programmes 2013.1.00351.S and 2015.1.00258.S (PI A.~Schruba), to be presented in A.~Schruba et~al.\ (in preparation) and first used in \citet{kruijssen19}. The observations were performed using the \mbox{12-m} main array, as well as the \mbox{7-m} array and total power antennas of the ALMA Compact Array, covering galactic radii out to $4.8$~kpc. The resulting data have angular resolution of $2.1\arcsec$ ($\sim$20~pc) and sensitivity of $0.1$~K per 2~\kms channel. Integrating over 10~\kms,\ this noise level translates to a 5$\sigma$ point source sensitivity of 5$\sigma\rm(M_{H_{2}})\approx 4.3\times10^3~M_{\odot}$ (assuming the $\rm\alpha_{CO}$ from Table~\ref{tab:input}). We retrieve \textit{Spitzer} MIPS 24$\rm\,\mu m$ image from the Local Volume Legacy (LVL) Survey \citep{dale09} and apply the same post-processing procedures as described above for IC\,342. We use the H$\alpha$ image presented in \citet{fasesi14}. This map is created from narrow-band H$\alpha$ data and nearby continuum available in the ESO data archive, and we use here the version kindly shared by Chris Faesi (private communication). The observations were carried out with the Wide Field Imager (WFI) on the MPG/ESO \mbox{2.2-m} telescope at La Silla observatory. Correction for Galactic extinction is applied using $A({\rm H\alpha}) = 0.027$~mag, obtained from $E(B{-}V)=0.01$~mag \citep{schlafly11} and an extinction curve with $R_{\rm V}=3.1$ \citep{fitzpatrick99}. We remove contamination of \NII lines by assuming an intensity ratio $I({\normalfont\textsc{N\,ii}})/I({\rm H}\alpha)=0.2$. The resolution of the map is $1.35\arcsec$. In our analysis, we consider emission from the field of view of the CO observations, and within $0-3$~kpc in galactic radius (beyond which the molecular gas surface density drops precipitously), and the outer boundary is visible in Figure~\ref{fig:map_rgb}.

\subsection{Homogenization of maps to common pixel grid}\label{subsec:regrid}
In order to apply our method, the gas and SFR tracer maps for a given galaxy need to share the same pixel grid. Therefore, for each galaxy, we regrid the map with a smaller pixel size to match the pixel grid of the map with larger pixel size. When the map that is being regridded has a better spatial resolution than the reference map, we first convolve the map with a Gaussian kernel to the resolution of the reference map before regridding to avoid introducing artifacts\footnote{When convolving the 24$\rm\,\mu m$ map, we have also tested using a more exact kernel from \citet{aniano11} and found that the use of a Gaussian kernel has a negligible impact on our results.}.  

\subsection{Construction of masks}\label{subsec:mask}

We use the small-scale variation of the gas-to-SFR flux ratios to constrain the evolutionary timeline of the molecular clouds (see Section~\ref{sec:method}). By definition, our measurements are flux-weighted averages (see \citealp{kruijssen18}), which implies that very bright peaks dominating a significant fraction of the total flux can bias our results. Therefore, we mask star-forming regions in some galaxies that are clear outliers in the luminosity function of SFR tracer peaks. Specifically, we first sort the peak fluxes (identified using \textsc{Clumpfind}; see Section~\ref{sec:method}) in descending order. We then look for a gap in the distribution by calculating the ratio of the flux between the $n^{\rm th}$ brightest and the next brightest peak in line, starting from the brightest peak. A gap is defined to exist when the $n^{\rm th}$ peak is more than twice as bright as the $(n{+}1)^{\rm th}$ peak. Whenever a gap is found, we mask all the peaks that are brighter than the $(n{+}1)^{\rm th}$ brightest peak. As a result, we mask three star-forming regions each in the LMC and in M33 before applying our analysis (green circles in Figure~\ref{fig:map_rgb}). These regions include 30 Doradus in the LMC and NGC\,604 in M33, which alone contribute more than 30\% of the 24$\rm\,\mu m$ emission of each galaxy. Note that we would be masking the same peaks unless we go down to a brightness difference of 50\% (rather than 100\%) when defining a gap in the luminosity function. In this case, we would be masking one to four more peaks each in IC\,342, M31, and NGC\,300. The impact of masking such bright regions on the resulting derived parameters is generally small when averaging over the entire galaxy, but becomes significant if a smaller fraction of the galaxy is considered (see \citealt{ward20} for the effect of 30 Doradus on the LMC and \citealt{chevance20} for the effect of the ``headlight cloud'' on the spiral galaxy NGC\,628, also see \citealt{herrera20}). We also check for bright regions that satisfy this condition in the CO emission maps, but found none. Finally, we also mask artifacts in the maps (purple circles in Figure~\ref{fig:map_rgb}). 

\section{Method}\label{sec:method}
We employ a statistical method (formalised in the \Heisenberg\ code) to constrain the evolutionary timeline of GMCs. This timeline can be decomposed into the cloud lifetime, the duration of the embedded phase of star formation (which continues until dispersal of molecular clouds), and the star formation tracer lifetime. The characteristic separation length between star-forming regions undergoing independent evolution is also constrained in our analysis. Here, we provide a summary of the methodology and the main input parameters. We refer the reader to \citet{kruijssen14} for a detailed explanation of the method, to \citet{kruijssen18} for the presentation and validation of the \Heisenberg\ code, as well as the full list of input parameters, and to \citet{chevance20} for a general application of the method to nine nearby star-forming galaxies. The accuracy of the method has been demonstrated in \citet{kruijssen18} using simulated galaxies, and has since been confirmed through extensive observational and numerical testing \citep{kruijssen19,haydon20_ext,ward20_hI}.

\begin{table*}
\begin{center}
\caption{Input parameters of the analysis using 24$\rm\,\mu m$ as SFR tracers for each galaxy. Other parameters not mentioned here are the same as in our previous analysis using H$\alpha$ as an SFR tracer.\label{tab:input}}
\begin{threeparttable}
\begin{tabular}{lccccccl}
\hline
Quantity &	 IC\,342 &	 LMC &	 M31 &	 M33 	& M51&	 NGC\,300 & Description\\
\hline					
$l_{\rm ap, min}$ [pc] &	116	& 25	&52&	65&	90	&60 & Minimum aperture size to convolve the input maps to\\
$l_{\rm ap, max}$ [pc] &	3000 &	2000&	3000&	2500&	3000&	2560 & Maximum aperture size to convolve the input maps to\\
$N_{\rm ap}$&	15&	15&	15&	15&	15&	15 & Number of aperture sizes used to create array of logarithmically-spaced \\ &&&&&&& aperture size in the range ($l_{\rm ap, min}$, $l_{\rm ap, max}$) \\
$\rm N_{pix, min}$& 	20&	10&	20&	20&	10&	100 & Minimum number of pixels for a valid peak\\
$\Delta\rm log_{10}\mathcal{F}_{\rm CO}$& 	2.0&	2.5&	1.3&	2.2&	2.5&	2.0 & Logarithmic range below flux maximum covered by flux contour levels \\ &&&&&&& for molecular gas peak identification \\
$\delta\rm log_{10}\mathcal{F}_{\rm CO}$ &	0.05&	0.15&	0.02&	0.10&	0.05&	0.10 & Logarithmic interval between flux contour levels for molecular gas peak \\ &&&&&&& identification \\
$\Delta\rm log_{10}\mathcal{F}_{\rm 24\,\mu m}$ &	3.8	&2.8&	2.3	&3.0&	4.0&	2.0 & Logarithmic range below flux maximum covered by flux contour levels\\ &&&&&&&  for SFR tracer peak identification\\ 
$\delta\rm log_{10}\mathcal{F}_{\rm 24\,\mu m}$& 	0.05&	0.05&	0.05	&0.10&	0.05&	0.10 & Logarithmic interval between flux contour levels for SFR tracer peak \\ &&&&&&& identification \\
$t_{\rm ref}$ [Myr] &	20.0&	11.1&	14.0&	14.5&	30.5&	10.8 & Reference timescale spanned by molecular gas tracer\\
$t_{\rm ref, errmin}$ [Myr] &	2.3&	1.7&	1.9&	1.5	&4.8&	1.7 &  Downwards uncertainty on reference timescale\\
$t_{\rm ref, errmax}$ [Myr]& 	2.0&	1.6&	2.1&	1.6&	9.2&	2.1&  Upwards uncertainty on reference timescale \\
SFR [$\rm M_{\odot}\pyr$]&	0.97&	0.12&	0.041&	0.18&	1.63&	0.063 & Total SFR in the analysed area \\
$\sigma$(SFR) [$\rm M_{\odot}\pyr$] &	0.19&	0.03&	0.008&	0.04&	0.32&	0.013 & Uncertainty of the total SFR \\
$\rm log_{10}~\alpha_{CO}$ & 	0.65&	0.83&	0.69&	0.81&	0.59&	0.82 & Logarithm of \coone-to-$\rm H_{2}$ conversion factor \\
$\sigma_{\rm rel}(\rm \alpha_{CO})$& 	0.5&	0.5&	0.5&	0.5&	0.5&	0.5 & Relative uncertainty of $\rm\alpha_{CO}$\\
$n_{\lambda}$&	13&	7&	10&	10&	16&	8 & Characteristic width for the Gaussian filter used to remove diffuse\\ &&&&&&&  emission in Fourier space \\
\hline
\end{tabular}
\end{threeparttable}
\end{center}
\end{table*}

\subsection{Description of the analysis method}
Galaxies are composed of numerous GMCs and star-forming regions. The fundamental concept of our method is that such regions are independently undergoing their evolution, from molecular clouds to the formation of stars. These evolutionary phases are observed using gas (e.g. CO) and SFR tracers (e.g. H$\alpha$ or 24$\rm\,\mu m$). We define the duration of each phase based on the visibility timescale of the tracers used. The timescale during which a gas emission tracer and an SFR tracer co-exist corresponds to the duration of massive star formation plus the time it takes to disrupt its natal molecular gas by stellar feedback (i.e. the feedback timescale, $t_{\rm fb}$). In the following, the cloud lifetime will be denoted as $t_{\rm CO}$, the star formation tracer lifetime as $t_{\rm H \alpha}$ or $t_{\rm 24\,\mu m}$, and the feedback timescale as $t_{\rm fb,\,H\alpha}$ or $t_{\rm fb,\,24 \,\mu m}$ depending on the SFR tracer used. 

During the initial phase of cloud evolution, a given independent region is only visible in the molecular gas tracer. As the cloud collapses and starts forming stars, the region becomes visible both in the gas and SFR tracers. Eventually, the remaining molecular gas is dispersed by stellar feedback and the region is only visible in the SFR tracers. Locally, the gas-to-SFR flux ratio therefore decreases with time during the evolution of a cloud. Observationally, when focusing on a non-star-forming GMC, a higher gas-to-SFR flux ratio is measured compared to the large-scale ($\sim$1~kpc) average gas-to-SFR flux ratio. By contrast, when focusing on a young star-forming region, where most of the molecular gas has been dispersed, a lower gas-to SFR flux ratio is measured. The deviations of the small-scale gas-to-SFR flux ratio compared to the large-scale average, as a function of spatial scale, can be directly related to the duration of the different phases of the GMC lifecycle \citep{kruijssen14, kruijssen18}.

In practice, we first identify peaks in the gas tracer and SFR tracer emission maps. We then convolve both maps into a range of $N_{\rm ap}$ spatial resolutions spanning from $l_{\rm ap, min}$ to $l_{\rm ap, max}$ (see Table~\ref{tab:input}). The minimum aperture size ($l_{\rm ap, min}$) is set to a value that is close to the size of the major axis of the deprojected beam of the coarsest resolution between the two maps, whereas the maximum aperture size ($l_{\rm ap, max}$) covers most of the galaxy. For each convolved map, apertures with the size of the corresponding resolution are placed on the identified gas and SFR tracer peaks. We then measure the gas and SFR tracer flux enclosed in these apertures to obtain the gas-to-SFR flux ratios as a function of aperture size. By fitting an analytical model describing the gas-to-SFR flux ratio as a function of the aperture size and the underlying evolutionary timescales, we obtain a direct measurement of these timescales. This can be understood with an idealised example. For a tracer that is longer lived, more peaks are typically identified, covering a larger fraction of the galaxy when small apertures are centred on them, compared to the shorter-lived tracer. The measured flux ratio is therefore closer to the galactic average value for a longer-lived tracer than a shorter-lived one.

We fit the analytical model derived by \citet{kruijssen18} to the measured flux ratios in order to constrain the relative duration of the different phases of the molecular cloud and star-forming region lifecycle, as well as the typical separation length between independent regions ($\lambda$). The absolute duration of the different phases is then obtained by scaling the relative duration of timescales with a reference timescale ($t_{\rm ref}$). In our previous analyses using CO and H$\alpha$ observations \citep{kruijssen19, chevance20, chevance20_fb, hygatePhD, ward20}, we used the duration of the isolated H$\alpha$ emitting phase ($t_{\rm ref}=t_{\rm H\alpha}-t_{\rm fb,\,H\alpha}$), calibrated by \citet{haydon20, haydon20_ext} using the stellar population synthesis model \textsc{SLUG2} \citep{dasilva12,dasilva14, krumholz15}, as the reference timescale. Here, in order to obtain absolute values when applying our analysis to CO and 24$\rm\,\mu m$ maps, we first apply the method to CO and H$\alpha$ observations. This is to obtain the cloud lifetime ($t_{\rm CO}$) and its upward and downward uncertainties ($t_{\rm CO,\,errmin}$ and $t_{\rm CO,\,errmax}$; see Table~\ref{tab:results}), which are adopted as the reference timescale ($t_{\rm ref}$) and its uncertainties ($t_{\rm ref,\,errmin}$ and $t_{\rm ref,\,errmax}$) in the analysis with CO and 24$\rm\,\mu m$ observations. The fitted model is thus described by three independent and non-degenerate quantities ($t_{\rm 24\mu m}$, $t_{\rm fb,\,24\mu m}$, and $\lambda$). The best-fitting values are then obtained by minimising the reduced-$\chi^{2}$ over these three quantities. The uncertainties of each parameter are propagated consistently throughout the analysis.

The presence of diffuse emission can bias our measurements by adding a large-scale component that is not associated with the identified peaks. This large-scale emission potentially includes diffuse emission originating from sources related to the recent massive star formation, such as low mass molecular clouds, low luminosity \HII regions, and ionizing photons that have escaped from \HII regions (e.g. \citealt{wood10}; F.~Belfiore et~al.\ in preparation). It may also originate from other mechanisms not related to recent massive star formation, for example, diffuse molecular gas, infrared emission powered by stars of intermediate age, and diffuse ionized gas created by shocks \citep{martin97, leroy12}. We remove such diffuse emission in both gas and SFR tracer emission maps iteratively, using the method presented in \citet{hygate19}, which makes the derived timescales sensitive to only the massive/\linebreak[0]{}luminous molecular clouds and young stellar populations. This method filters emission on spatial scales larger than $n_{\lambda}$ times the typical distance between regions $\lambda$ (as measured from the \Heisenberg\ code) using a Gaussian high-pass filter in Fourier space. For each galaxy, we adopt the smallest possible value for $n_{\lambda}$, while ensuring the flux loss from the compact emission to be less than 10\% (also following \citealt{chevance20,hygatePhD}; see Table~\ref{tab:input}). We do not adopt a fixed filtering scale because we want to maximise the removal of diffuse emission, while minimising the impact of the filtering on the compact regions. The influence of $n_{\lambda}$ on the derived timescales is fully described in \citet{hygate19} and \citet{hygatePhD}. In summary, the choice of $n_{\lambda}$ does not significantly change the best-fitting model parameters, as long as the adopted $n_{\lambda}$ is smaller than 30 and the flux loss from the compact emission is less than 10\%. After the diffuse emission is filtered out, a noise mask with a threshold at twice the standard deviation noise level of the emission map is applied. We repeat this process until the convergence condition is reached, which is when the change of the measured value of $\lambda$ is less than 5\% over three consecutive iterations.

\subsection{Input parameters}\label{subsec:input}

Unless otherwise noted here, we adopt the same parameters as for previous analyses using H$\alpha$ as an SFR tracer (see Appendix~\ref{sec:app_tco} for IC\,342 and M31, \citealt{ward20} for the LMC, \citealt{hygatePhD} for M33, \citealt{chevance20} for M51, and \citealt{kruijssen19} for NGC\,300). The parameters not mentioned here include distance, inclination, position angle (see Table~\ref{tab:obs}), as well as parameters related to the fitting process and error propagation, for which default values are adopted as listed in \citet{kruijssen18}. We use \textsc{Clumpfind} \citep{williams94} to identify gas and SFR tracer peaks in each map. This algorithm finds peaks by drawing closed contours for a set of flux levels, within a given flux range ($\rm{\Delta}log_{10}\mathcal{F}$) below the maximum flux level, with an interval of $\rm{\delta}log_{10}\mathcal{F}$ between flux levels. The adopted values for our sample are summarised in Table~\ref{tab:input}. Moreover, to avoid identifying point sources that are likely to be foreground stars that were not masked during the image reduction process or externally illuminated starless dust clumps (see Section~\ref{subsec:eff_evolved}), we only accept peaks that contain more than $N_{\rm pix,min}$ pixels. The area of $N_{\rm pix,min}$ pixels equals $0.2 - 3.5$ times the coarsest beam size. We note that our choices of $\rm{\Delta}log_{10}\mathcal{F}$, $\rm{\delta}log_{10}\mathcal{F}$, $N_{\rm pix,min}$, $l_{\rm ap,max}$, and $N_{\rm ap}$ do not affect our measurements significantly as long as peaks that are obviously visible in the emission maps have been identified \citep{kruijssen18}. As explained above, $t_{\rm CO}$ and its uncertainties determined from our analysis with CO and H$\alpha$ are used to define the reference timescales ($t_{\rm ref}$). However, since we additionally mask some of the bright star-forming regions for the reasons explained in Section~\ref{subsec:mask}, we re-run the same analysis using H$\alpha$ as an SFR tracer with updated masks. Our measurements using H$\alpha$ as an SFR tracer are listed in Table~\ref{tab:results}, and are in very good agreement with (or identical to) the previously published results. For the analysis with CO and 24$\rm\,\mu m$, $t_{\rm ref}~(=t_{\rm CO})$ includes the feedback phase. The \coone-to-$\rm H_{2}$ conversion factor ($\rm \alpha_{CO}$; including the contribution from heavy elements) is adopted from \citet{bolatto13}, expressed as 
\begin{equation}
    \alpha_{\rm CO} = \left[2.9~\Msun~({\rm K}~\kms~\pc^2)^{-1}\right] \times \exp\left(\frac{0.4\,{\rm Z}_{\sun}}{Z}\right)~.
\end{equation}\label{eq:conv_gas}
We adopt the metallicity-dependent part, but not the surface density dependence from \citet{bolatto13}. For simplicity, we adopt a constant $\rm\alpha_{CO}$ value for each galaxy and a conservative uncertainty of 50\%. In addition, for M33, which is the only galaxy with \cotwo data, we adopt a fixed ratio of $\cotwo/\coone=0.8$ \citep{gratier10}. Finally, we derive the total SFR for the analysed area by combining 24$\rm\,\mu m$ and H$\alpha$ emission maps and using the conversion factor from \citet{calzetti07}, expressed as
 \begin{equation}
    \rm SFR(\rm\Msun~yr^{-1}) = ~5.3 \times 10^{-42}\left[\textit{L}(H\alpha)+0.031\textit{L}(24\,\mu m)\right]~,
    \label{eq:conv_sfr}
\end{equation}
where the luminosities have units of $\rm erg\,s^{-1}$ and \textit{L}($\rm 24\,\mu m$) is expressed as $\rm \nu\textit{L}(\nu)$. We assume a typical uncertainty of 20\% for the derived SFR. These conversion factors are only used to derive additional physical quantities such as the molecular gas surface density and the integrated star formation efficiency. We note that the exact values of these conversion factors, unless they vary spatially, do not affect our measurements of primary quantities, which are the durations of the successive phases of cloud evolution and star formation, nor do they affect the region separation length between independent regions.

\begin{figure*}
\includegraphics[width=\linewidth]{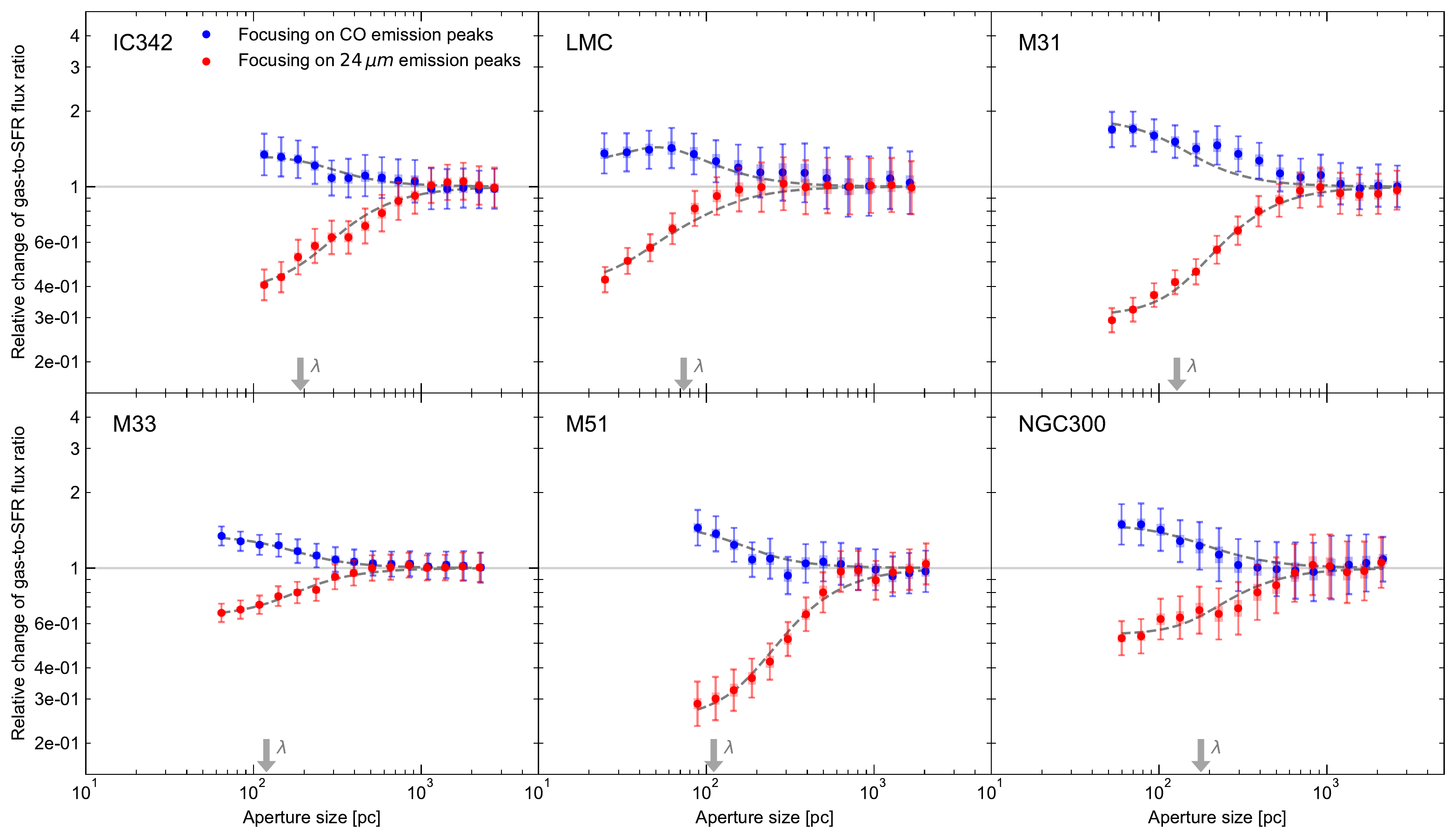}
\caption{Relative change of the gas-to-SFR (CO-to-24$\rm\,\mu m$) flux ratio compared to the galactic average as a function of the size of apertures placed on CO (blue) and 24$\rm\,\mu m$ (red) emission peaks. The error bars indicate 1$\sigma$ uncertainty on each individual data point whereas the shaded area is an effective 1$\sigma$ uncertainty taking into account the covariance between data points. The galactic average is shown as the solid horizontal line and the dashed line indicates our best-fitting model. The constrained region separation length ($\lambda$) is indicated in each panel with the downward arrow and other constrained best-fitting parameters ($t_{\rm fb,\,24\,\mu m}$, and $t_{\rm 24\,\mu m}$) are listed in Table~\ref{tab:results}. }
\label{fig:tuningfork}
\end{figure*}

\section{Results}\label{sec:result}

\subsection{The molecular cloud lifecycle}

\begin{table*}
\begin{center}
{\def\arraystretch{1.5}
\caption{Physical quantities constrained using the method described in Section~\ref{sec:method}, describing the evolution of molecular clouds to exposed or embedded stellar populations traced by H$\alpha$ and 24$\rm\,\mu m$, respectively. Following the notation throughout this paper, $t_{\rm CO}$ is the cloud lifetime, $t_{\rm fb,\,H\alpha}$ and $t_{\rm fb,\,24\,\mu m}$ are the duration of the partially exposed and embedded star-forming phase, respectively, and \tHa\ and \tir\ are the duration of H$\alpha$ and 24$\rm\,\mu m$ emitting phase, respectively. The region separation length ($\lambda$) measured with different SFR tracers, feedback velocity ($v_{\rm fb}$), and star-formation efficiency (\esf)\ are also listed.}
\label{tab:results}
\begin{threeparttable}
\begin{tabular}{lcccccccccc}
\hline
&\multicolumn{4}{c}{CO vs \Ha} && \multicolumn{3}{c}{CO vs 24$\rm\,\mu m$} \\
 \cline{2-5}\cline{7-9}
Galaxy & \tCO\ & $t_{\rm fb,\,H\alpha}$ & \tHa\ & $\lambda$    && $t_{\rm fb,\,24\,\mu m}$ & \tir\ & $\lambda$  &  \vfb\ &\esf\  \\
 & [Myr] & [Myr] & [Myr] & [pc]  && [Myr] & [Myr] & [pc] & [km s$^{-1}$]& [per~cent] \\
\hline
IC\,342
& $20.0_{-2.3}^{+2.0}$& $2.2_{-0.5}^{+0.4}$& $6.4_{-0.6}^{+0.5}$& $120_{-10}^{+10}$&& $5.2_{-2.3}^{+1.5}$& $7.9_{-2.2}^{+1.8}$& $190_{-62}^{+59}$& $14.3_{-1.8}^{+4.0}$& $1.9_{-0.8}^{+1.4}$\\
LMC
& $11.1_{-1.7}^{+1.6}$& $1.2_{-0.2}^{+0.2}$& $5.8_{-0.4}^{+0.4}$& $71_{-8}^{+13}$&& $5.0_{-2.0}^{+1.6}$& $13.6_{-4.8}^{+3.7}$& $73_{-26}^{+38}$& $10.0_{-1.7}^{+2.1}$& $6.8_{-3.0}^{+4.9}$\\
M31
& $14.0_{-1.9}^{+2.1}$& $1.1_{-0.2}^{+0.3}$& $5.5_{-0.3}^{+0.4}$& $181_{-19}^{+28}$&& $2.4_{-0.8}^{+1.4}$& $4.2_{-0.7}^{+1.5}$& $128_{-23}^{+97}$& $29.5_{-5.3}^{+6.9}$& $0.7_{-0.2}^{+0.2}$\\
M33
& $14.5_{-1.5}^{+1.6}$& $3.3_{-0.5}^{+0.6}$& $7.9_{-0.6}^{+0.7}$& $155_{-24}^{+30}$&& $6.8_{-2.0}^{+2.1}$& $11.9_{-2.1}^{+2.9}$& $119_{-35}^{+60}$& $10.3_{-1.3}^{+1.5}$& $3.5_{-1.5}^{+2.5}$\\
M51
& $30.7_{-4.9}^{+8.7}$& $4.7_{-1.1}^{+2.0}$& $8.9_{-1.2}^{+2.0}$& $140_{-17}^{+25}$&& $<4.0^{a}$& $3.6_{-0.9}^{+1.2}$& $<136^{a}$& $7.9_{-2.1}^{+2.0}$& $3.3_{-1.4}^{+2.9}$\\
NGC\,300
& $10.8_{-1.6}^{+2.2}$& $1.5_{-0.2}^{+0.2}$& $6.1_{-0.2}^{+0.2}$& $104_{-18}^{+22}$&& $4.9_{-1.9}^{+1.2}$& $7.9_{-2.1}^{+1.5}$& $178_{-75}^{+125}$& $9.4_{-0.7}^{+0.8}$& $3.3_{-1.4}^{+2.6}$\\
\hline
\end{tabular}
\begin{tablenotes}
\item[\textit{a}]Only a $1\sigma$ upper limit can be derived for not satisfying (ii) and (viii) in Section~\ref{subsec:robust_kl18}.
\end{tablenotes}
\end{threeparttable}
}
\end{center}
\end{table*}

\begin{figure*}
\includegraphics[width=\linewidth]{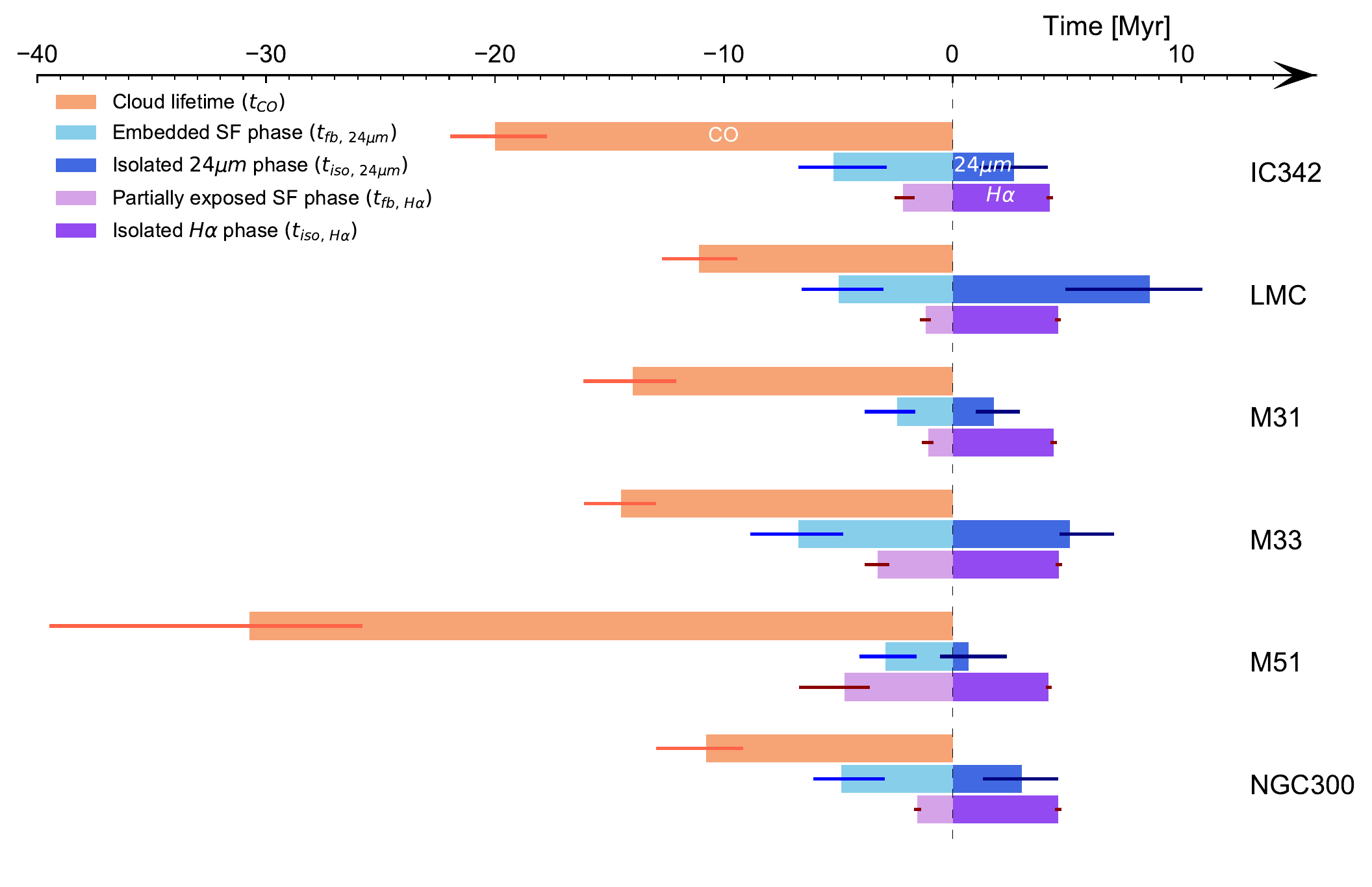}
\caption{Timeline describing the evolution from molecular clouds to the embedded star-forming phase and then finally to exposed young stellar regions. The time during which CO is visible ($=t_{\rm CO}$) is indicated in orange, the time during which 24$\rm\,\mu m$ and H$\alpha$ are visible without CO are shown respectively in dark blue and dark purple. The timescales for the feedback phase, during which both CO and SFR tracer emissions are observed co-spatially are shown in light blue (for 24$\rm\,\mu m$) and light purple (for H$\alpha$). The corresponding 1$\sigma$ error bars are also indicated. We note that for M51 the feedback timescale constrained using 24$\rm\,\mu m$ ($t_{\rm fb,\,24\,\mu m}$) could have been biased by deconvolution artifacts (see Section~\ref{subsubsec:tfb}). }
\label{fig:timeline}
\end{figure*}

Here, we present our results from the application of our method to the maps of CO and 24$\rm\,\mu m$ emission presented in Section~\ref{sec:data}, as tracers of the molecular gas and the SFR for six nearby galaxies. Figure~\ref{fig:tuningfork} shows the gas-to-SFR flux ratios measured around gas and SFR tracer peaks, as a function of aperture size, together with our best-fitting model for each galaxy. Going towards smaller aperture sizes (from $\sim$1 kpc to $\sim$50~pc), the measured flux ratios for both branches increasingly deviate from the galactic average, illustrating the spatial de-correlation between the gas and SFR tracer emission peaks. Table~\ref{tab:results} summarises the constrained best-fitting values from applying our analysis to the H$\alpha$ and CO maps, as well as to the 24$\rm\,\mu m$ and CO maps. The first experiment allows us to measure $t_{\rm CO}$, which is then used as the reference timescale for the second experiment (see Section~\ref{sec:method}). Table~\ref{tab:results} also lists other physical quantities which can be derived from our measurements, such as the feedback outflow velocity ($v_{\rm fb}$; see Section~\ref{subsubsec:vfb}) and the integrated cloud-scale star formation efficiency ($ \epsilon_{\rm sf}$; see Section~\ref{subsubsec:esf}). In Figure~\ref{fig:timeline}, we show an illustration of the evolutionary timelines of molecular clouds and star-forming regions in our galaxy sample. GMCs initially emit only in CO, then in 24\mic\ after the onset of star formation and finally in H$\alpha$ when the star-forming regions become (partially) exposed. 

\subsubsection{Feedback timescale}\label{subsubsec:tfb}
The use of 24$\rm\,\mu m$ emission enables us to take the heavily obscured phase of star formation into account, which cannot be done with the analysis of only CO and H$\rm\alpha$ emission. The duration of the embedded phase of massive star formation (i.e. feedback timescale; $t_{\rm fb,\,24\,\mu m}$), which continues until disruption of molecular clouds, is measured to be $2-7$~Myr in our sample of galaxies. Our measurements suggest that molecular clouds spend $17-47\%$ of their lifetime with massive stars embedded. For almost all of the galaxies in our sample (except M51), the measured $t_{\rm fb,\,24\,\mu m}$ is $1-4$~Myr longer than the one obtained using H$\rm \alpha$ emission ($t_{\rm fb,\,H\alpha}$; see Table~\ref{tab:results}). This is expected, as 24$\rm\,\mu m$ is already detected during the heavily obscured phase of star formation, making it visible for a longer duration than H$\alpha$, which is only detectable when massive stars have formed and surrounding gas and dust have been partially cleared out. By contrast, in the particular case of M51, we find $t_{\rm fb,\,24\,\mu m}$ to be shorter than (or comparable within 1$\sigma$ uncertainty to) the feedback timescale obtained using H$\alpha$. We suspect that such a potentially unphysical measurement could be due to artifacts in the 24$\rm\,\mu m$ map of M51 introduced by the deconvolution algorithm used to create the high resolution map \citep{dumas11}. These artifacts are clearly visible as dark rings around bright peaks in the spiral arms of the galaxy (see Figure~\ref{fig:map_rgb} and Appendix~\ref{sec:app_map}), and may (or may not) make the inferred timescales less accurate as discussed in Section~\ref{subsec:datasets}.

The measured durations between the onset of embedded star formation and molecular cloud disruption ($2-7$~Myr), are comparable to the time it takes for the first supernova to explode. This is about $4$~Myr for a fully populated stellar initial mass function \citep{leitherer14} and can be up to $20$~Myr when the initial mass function is stochastically sampled for the typical stellar region masses considered here (with almost no dependence on metallicity; \citealp{chevance20_fb}). The measured short feedback timescales indicate that pre-supernova feedback such as photoionization and stellar winds are mostly responsible for the dispersal of molecular clouds. Our measurements of the feedback timescale show a good agreement with the typical age of star clusters when they stop being associated with their natal GMCs both in the Milky Way and nearby galaxies ($2-7$~Myr; \citealt{lada03, whitmore14, hollyhead15, corbelli17, grasha18, grasha19}). This is further discussed in Section~\ref{subsec:comparison}. Radiation magnetohydrodynamic simulations of GMCs by \citet{kim20} suggest a similar duration of the star formation and feedback timescales ($4-8$~Myr). The measured duration for embedded star formation is somewhat shorter than the age spread of star clusters in the LMC ($7-12$~Myr) measured by \citet{efremov98}, on the scale of the mean radius of SFR tracer peaks (${\sim}10-50$~pc). This is expected as the actual size of a star-forming region is necessarily smaller than the size of a SFR tracer peak, which is limited by the spatial resolution of our maps. The agreement gets better if we only consider the age spreads measured among young stellar clusters ($1-4$~Myr; from star clusters with ages of $1-10$~Myr).

\subsubsection{Duration of the 24\,$\mu m$ emitting phase}
Across all galaxies in our sample, we find that the 24$\rm\,\mu m$ emission phase lasts for $4-14$~Myr. For M33, we note that our measurement of $t_{\rm24\,\mu m}$ is in very good agreement with the one from \citet{corbelli17}, where 24$\rm\,\mu m$ emission is found to last for $\sim$10~Myr after the onset of massive star formation by applying a cloud classification method to M33. The ratio between duration of visibility for 24$\rm\,\mu m$ and H$\alpha$ emissions ($t_{\rm24\,\mu m}/t_{\rm H\alpha}$) ranges from $0.4$ to $2.3$.

The 24$\rm\,\mu m$ emission does not originate only from embedded young massive stars, but also from late-type B~stars and the interstellar radiation field, which make a non-negligible contribution to the dust heating \citep{draine07, verley09}. The contribution of the sources not related to recent local massive star formation to the 24$\rm\,\mu m$ emission is more homogeneously spread in the galaxy compared to the more clustered young stellar population \citep[e.g.][]{dale07, leroy12} and results in an additional diffuse component of the 24$\rm\,\mu m$ emission on large scales. We separate this diffuse emission from the compact emission of young stellar regions by applying the filtering process described in Section~\ref{sec:method}. The fact that in most galaxies the end of the 24$\rm\,\mu m$ emission phase is before or similar to the end of the H$\alpha$ emission phase (see Figure~\ref{fig:timeline}) shows that this procedure effectively removed contamination from stellar populations not related to recent massive star formation. In Section~\ref{subsec:eff_evolved}, we discuss in more detail the effects of starless dust clumps illuminated by external radiation, and late-type B~stars still preferentially located near their birth sites, which might not be removed by our filtering process.

We note that, while the duration of the isolated H$\alpha$ emitting phase ($t_{\rm iso,H\alpha} = t_{\rm H\alpha} - t_{\rm fb,\,H\alpha}$) is almost constant in all the galaxies in our sample by construction \citep{haydon20}, the duration of isolated 24$\rm\,\mu m$ emission ($t_{\rm iso, 24\,\mu m} = t_{\rm 24\,\mu m} - t_{\rm fb,\,24\,\mu m}$) appears to vary across the sample ranging from $\rm\sim2~Myr$ (excluding M51) to the end of the H$\alpha$ emitting phase ($8.6_{-3.7}^{+2.3}$~Myr). This isolated phase originates from stochastic heating of small dust grains in the CO dark clouds. We find this phase to be shorter for star-forming regions in more metal-rich galaxies (see Figure~\ref{fig:timeline} and \ref{fig:env}). This is discussed in more detail in Section~\ref{subsec:envdep}.

\subsubsection{Region separation length}\label{subsubsec:lambda}
As visible in Figure~\ref{fig:tuningfork}, gas and SFR tracer peaks are spatially decorrelated on small spatial scales, revealing that galaxies are made of regions that are independently undergoing evolution from molecular gas to stars. The spatial scale at which the gas-to-SFR ratio diverges from the galactic average (Figure~\ref{fig:tuningfork}) is linked to the typical distance $\lambda$ between independent regions. We find that $\lambda$ ranges from $70$~pc to $190$~pc for the galaxies in our sample when considering the 24$\rm\,\mu m$ and CO emission maps. For M51, we do not sufficiently resolve the region separation length and are only able to obtain an upper limit of $\lambda$ (see Section~\ref{subsec:robust_kl18}). For the other galaxies, we find that $\lambda$ derived using 24$\rm\,\mu m$ maps and H$\alpha$ maps agree to within the formal uncertainties. 

While the physical mechanisms that set the region separation length remain debated, a similarity between the region separation length and the gas disc scale height has been reported by \citet{kruijssen19} in NGC\,300, suggesting that the depressurisation of \HII regions along the direction perpendicular to the galactic disc might be responsible for this characteristic length. Furthermore, the measured values of $\lambda$ are comparable to the thickness of the vertical distribution of star-forming regions undergoing the earliest stages of evolution ($150-200$~pc) observed using \textit{Spitzer} IRAC 8$\rm\,\mu m$ maps of edge-on spiral galaxies (NGC\,891 and IC\,5052; \citealp{elmegreen20}), to the thickness of the molecular disc measured with CO emission \citep{scoville93, yim14, heyer15, patra20, yim20}, and also to the amplitude of the oscillation seen in the Radcliffe Wave recently discovered in the vicinity of the Sun \citep{alves20}. Finally, the region separation length roughly coincides with the spatial wavelength of velocity corrugations in NGC\,4321 \citep{henshaw20}, indicating that it matches the scale on which the molecular interstellar medium is reorganised by cloud-scale matter flows. Further investigation is needed to verify quantitatively whether the correlation observed by \citet{kruijssen19} holds more generally in nearby galaxies.

\subsubsection{Star formation efficiency}
\label{subsubsec:esf}
The SFR surface density ($\Sigma_{\rm SFR}$) corresponds to the mass of newly formed stars inferred for a given SFR tracer, divided by that SFR tracer's emission timescale. Similarly, the rate of molecular gas formation can be expressed as $\Sigma_{\rm gas}/t_{\rm CO}$, where $\Sigma_{\rm gas}$ is the surface density of molecular gas and the $t_{\rm CO}$ is the timescale over which molecular gas assembles and form stars. By dividing these two rates, the time-averaged star formation efficiency per star-forming event ($\epsilon_{\rm sf}$) can be computed as:
\begin{equation}
\label{eq:esf}
\epsilon_{\rm sf} = \frac{t_{\rm CO} \Sigma_{\rm SFR}}{\Sigma_{\rm gas}}~.
\end{equation}
When calculating $\Sigma_{\rm gas}$, we only consider the compact CO emission, after the filtering of diffuse emission (see Section~\ref{sec:method}), which is also consistent with the flux we use to determine $t_{\rm CO}$. By doing this, we selectively include the CO emission  that participates in the massive star formation process while excluding emission that is likely to originate from diffuse molecular gas and small clouds. The filtering process removes 10\% to 50\% of the CO emission from the unfiltered maps. However, $\Sigma_{\rm SFR}$ is calculated using the total SFR, obtained by combining \Ha\ and 24$\rm\,\mu m$ emission to account for the effect of internal extinction (see Section~\ref{sec:data}). We note that our assumption implies that we attribute all of the diffuse emission in SFR tracer maps to recent massive star formation (e.g., leakage of ionizing photons from \HII regions). This ignores the fact that diffuse emission may also originate from mechanisms that are not related to recent massive star formation, such as diffuse ionized gas created by shocks and evolved post-asymptotic giant branch stars, as well as infrared emission powered by older stellar populations, which are known to have a relatively minor contribution to the dust heating \citep{nersesian20}. Under these conditions, we measure a low star formation efficiency per star-forming event in our sample of galaxies with $\epsilon_{\rm sf} = 0.7-6.8\%$.  This is consistent with previous measurements in these galaxies using H$\alpha$ as a tracer of recent star formation and other wavelengths such as \textit{GALEX} FUV and \textit{WISE} 22$\rm\,\mu m$ \citep{leroy12, leroy19} to estimate the global SFR \citep{chevance20, hygatePhD, ward20}. We note that for NGC\,300, we find $\epsilon_{\rm sf}$ to be slightly higher (but compatible within 1$\sigma$ uncertainty) than that measured in \citet{kruijssen19}. The difference is because \citet{kruijssen19} only considered H$\alpha$ emission when calculating the global SFR.

We also compare our measurements for $\epsilon_{\rm sf}$ to the fraction of gas converted into stars per gravitational free-fall time, which is expressed as $\epsilon_{\rm ff} = t_{\rm ff} \Sigma_{\rm SFR}/\Sigma_{\rm gas}$ and measured by \citealt{leroy17}, \citealt{utomo18} and \citet{schruba19} for most of the galaxies in our sample. We find that our measurements for the LMC, M31, M33, and NGC\,300 are somewhat ($\rm\leq4.0\%$) higher than the star formation efficiency per free-fall time ($\epsilon_{\rm ff}$) measured by \citet{schruba19}, which are $2.5$\%, $0.7$\%, $1.5$\%, and $1.2$\%, respectively. Because the cloud lifetime in these galaxies is similar to the free-fall timescale \citep{schruba19}, this difference is mostly due to the fact that we measure gas surface density from the diffuse emission filtered CO map, in order to calculate the fraction of compact clouds turning into stars. For M51, the difference becomes more significant ($\epsilon_{\rm ff}=0.3 {-}0.36\%$; \citealt{leroy17, utomo18}) and is because cloud lifetime is almost five times the free-fall timescale.

\subsubsection{Feedback velocity}\label{subsubsec:vfb}
As a result of the energetic feedback from young massive stars, CO emission becomes rapidly undetectable after the onset of star formation. This is most likely due to a phase and density change of the neighbouring medium through kinetic dispersal, ionization, and photodissociation. We combine the timescale over which molecular clouds are disrupted by feedback, $t_{\rm fb}$, with the characteristic size of the clouds detected in CO, $r_{\rm CO}$, to define the feedback velocity as $v_{\rm fb}=r_{\rm CO}/t_{\rm fb}$. The size of the cloud is measured in our method by fitting a Gaussian profile to the surface density contrast between the peak and the background, and ranges between $10{-}40$~pc for the sample of our galaxies. The velocity represents the speed of the kinetic removal of molecular gas or the phase transition front, depending on the nature of the dispersal mechanism. We use here $t_{\rm fb,\,H\alpha}$ to define $v_{\rm fb}$, assuming that disruption of the molecular gas starts with the young stellar population becoming partially exposed, whereas the 24$\rm\,\mu m$ is emitted even during the heavily obscured phase where the expansion of the \HII region has not yet begun and during which gas accretion onto the cloud potentially continues. 

The derived $v_{\rm fb}$ ranges between $9-30$~\kms, which is comparable to the measurements of expansion velocities of nearby \HII regions. For example, the expansion velocities of five \HII regions in NGC\,300 are measured to be in the range of $5-30$~\kms\ \citep{mcleod20}. Similar values are also found for \HII regions in the LMC \citep{naze01, naze02, mcleod19} and the Milky Way \citep{murray10, barnes20}. Numerical simulations of star-forming regions \citep[e.g.][]{kim18} also support this range of values.

\begin{figure*}
\includegraphics[width=16cm]{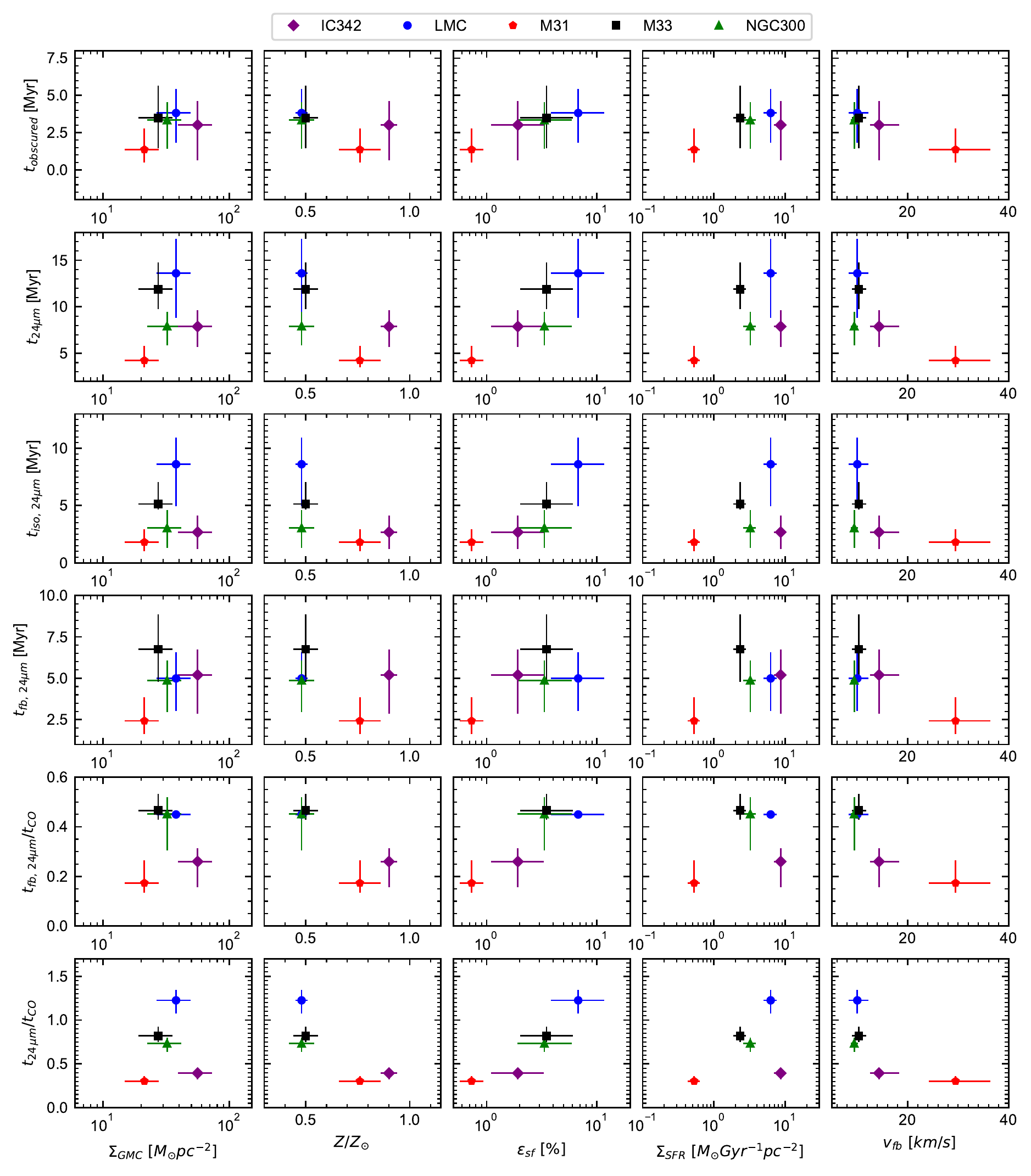}
\caption{In the top four rows, our measurements of the duration of the heavily obscured phase ($t_{\rm obscured}=t_{\rm fb,\,24\,\mu m}-t_{\rm fb,\,H\alpha}$), the 24$\rm\,\mu m$ emitting phase ($t_{\rm 24\,\mu m}$), the isolated 24$\rm\,\mu m$ emitting phase ($t_{\rm iso, 24\,\mu m}=t_{\rm 24\,\mu m}-t_{\rm fb,\,24\,\mu m}$), and the feedback phase ($t_{\rm fb,\,24\,\mu m}$) are shown as a function of galactic (or cloud-scale) properties, i.e.\ the mass-weighted mean molecular gas surface density of GMCs ($\Sigma_{\rm GMC}$) measured in \citet{schruba19} and \citet{schruba21_ic342}, metallicity ($Z/Z_{\odot}$), star formation efficiency ($\epsilon_{\rm sf}$), SFR surface density ($\Sigma_{\rm SFR}$) and the feedback velocity ($v_{\rm fb}$). The bottom two rows show the ratios of the feedback timescale and the 24$\rm\,\mu m$ emitting phase to the cloud lifetime ($t_{\rm fb,\,24\,\mu m}/t_{\rm CO}$ and $t_{\rm 24\,\mu m}/t_{\rm CO}$), as a function of the same galactic properties.} 
\label{fig:env}
\end{figure*}

\subsection{Heavily obscured phase of star formation}
\label{subsec:em}
In order to probe the earliest phase of star formation, which is only associated with 24$\rm\,\mu m$ emission and not with H$\alpha$ because of strong attenuation provided by the surrounding gas, we measure the duration of the heavily obscured phase of star formation as the difference between the feedback timescale for 24$\rm\,\mu m$ emission (i.e.\ the total duration of the embedded phase of star formation) and the one for H$\alpha$ emission (i.e.\ the duration of the partially exposed phase of star formation): $t_{\rm obscured}=t_{\rm fb,\,24\,\mu m}-t_{\rm fb,\,H\alpha}$. We omit M51 here because the insufficient resolution of the 24$\rm\,\mu m$ map only allows for the determination of an upper limit on $t_{\rm fb,\,24\,\mu m}$ (see Section~\ref{subsec:robust_kl18}) and the presence of deconvolution artifacts (see Section~\ref{subsec:datasets}) could have biased our measurements, especially for timescale-related quantities. However, we retain M51 in our sample at large, to show our (unsuccessful) attempt in applying the method to a 24$\rm\,\mu m$ emission map that has been created using the HiRes deconvolution algorithm \citep{backus05}. Across the rest of our galaxy sample, we find $t_{\rm obscured} = 3.0\pm0.9$~Myr, with a full range of $1.4-3.8$~Myr (see Figure~\ref{fig:env}). The measured duration is comparable to age estimates of heavily obscured star clusters in the Milky Way \citep[0.5-3~Myr; see][and references therein]{lada03}. A similar duration of the highly embedded star-forming phase has recently been reported by \citet{elmegreen19, elmegreen20}, where such a phase is suggested to last for $1-2$~Myr based on the mass measurement of star-forming cores in nearby spiral galaxies using 8$\rm\,\mu m$ emission. The measurement of $2.4$~Myr for the heavily obscured phase using a cloud classification method by \citet{corbelli17} also supports the values we obtain here (see more in Section~\ref{subsec:comparison}).

\subsection{Relation with environmental properties}
\label{subsec:envdep}
We now explore potential environmental dependences of the durations of the successive phases of cloud evolution and star formation. In Figure~\ref{fig:env}, our measurements of the durations of the heavily obscured phase ($t_{\rm obscured}$), the total 24$\rm\,\mu m$ emission phase ($t_{\rm 24\,\mu m}$), the isolated 24$\rm\,\mu m$ emission phase ($t_{\rm iso, 24\,\mu m}=t_{\rm 24\,\mu m}-t_{\rm fb,\,24\,\mu m}$), and the feedback phase ($t_{\rm fb,\,24\,\mu m}$) are shown in the top four rows as a function of local cloud-scale and galactic properties such as the mass-weighted mean molecular gas surface density of GMCs ($\Sigma_{\rm GMC}$; measured by \citealt{schruba21_ic342} for IC342 and \citealt{schruba19} for the other galaxies), metallicity (relative to solar metallicity), integrated star formation efficiency per star formation event ($\epsilon_{\rm sf}$), SFR surface density ($\Sigma_{\rm SFR}$) and the feedback velocity ($v_{\rm fb}$).\footnote{The measurement of $\Sigma_{\rm GMC}$ for IC\,342 is calculated excluding the five most massive clouds (with mass $>2\times10^7 M_{\odot}$), which contribute 13\% of the total mass contained in molecular clouds. Including these clouds would not change our conclusion.}

In order to look for correlations with these environmental properties, we use LINMIX \citep{kelly07}, a Bayesian method accounting for measurement errors in linear regression. We do not find any statistically significant trend, which is defined to exist when the correlation coefficient is positive or negative with 95\% probability. While this is not shown in Figure~\ref{fig:env}, we also found no statistically significant correlation with the galaxy-averaged molecular gas surface density ($\Sigma_{\rm gas}$). However, we note that $t_{\rm 24\,\mu m}$ and $t_{\rm iso, 24\,\mu m}$ appear to be somewhat shorter towards increasing metallicity and decreasing star formation efficiency. While it is difficult to distinguish what is driving this trend, as metallicity and star formation efficiency are correlated in our sample of galaxies, we suspect that metallicity could be the primary driver of this trend. Indeed, winds from massive stars become more energetic with increasing metallicity \citep{maeder92}, resulting in a faster dispersal of the surrounding gas and decay of $t_{\rm 24\,\mu m}$ emission. We note that one might expect to see longer (isolated) 24$\rm\,\mu m$ emitting timescale for metal-rich galaxies since the dust-to-gas ratio correlates with metallicity \citep{remy-ryuer14}. This is not what we find here, which can be explained by our use of relative changes of gas-to-SFR tracer flux ratio compared to the galactic average, instead of using absolute flux ratios, when constraining the timeline of GMC evolution. 

The relative fractions of $t_{\rm fb,\,24\,\mu m}$ and $t_{\rm 24\,\mu m}$ compared to $t_{\rm CO}$ are plotted against galactic properties in the bottom two rows of Figure~\ref{fig:env}. We again see an anticorrelation of these durations with metallicity and a correlation with star formation efficiency, while no statistically significant trend is found with respect to galactic properties using the same regression analysis \citep{kelly07}. Our results (bottom row) show that clouds in the LMC, M33 and NGC\,300 spend a larger fraction of their lifetime with embedded massive star formation ($\sim$40\%) compared to clouds in IC\,342 and M31 ($\sim$20\%), with an average of 36\% across our sample of galaxies.

\section{Discussion}\label{sec:discussion}

\subsection{Robustness of the results}\label{subsec:robust}
\subsubsection{Satisfaction of guidelines in \citet{kruijssen18}}\label{subsec:robust_kl18}
We verify here that our analysis satisfies the requirements listed in section~\mbox{4.4} of \citet{kruijssen18}. Satisfaction of these criteria indicates that the constrained parameters $t_{\rm24\,\mu m}$, $t_{\rm fb,\,24\,\mu m}$, and $\lambda$ are measured with an accuracy of at least 30\%. For the analysis using H$\alpha$ as an SFR tracer, we only check the accuracy for IC\,342 and M31, because the measurements for other galaxies have already been validated by previous studies \citep[see][]{kruijssen19, chevance20, hygatePhD, ward20}.
\begin{enumerate}
\item The duration of gas and stellar phases should always differ less than one order of magnitude. This condition is satisfied by $|\log_{10}(t_{\rm H\alpha}/t_{\rm CO})| \leq 0.5$ for IC\,342 and M31, and $|\log_{10}(t_{\rm 24 \,\mu m}/t_{\rm CO})| \leq 0.92$ for all the galaxies in our sample, where the difference between $t_{\rm 24 \,\mu m}$ and $t_{\rm CO}$ is the largest in M51 while those for other galaxies are $|\log_{10}(t_{\rm 24 \,\mu m}/t_{\rm CO})|~\leq~0.52$. 
\item For almost all of the galaxies (except M51), we measure $\lambda \geq 1.6 l_{\rm ap, min}$, which ensures that the region separation length is sufficiently resolved by our observations. For M51, we measure $\lambda = 1.2 l_{\rm ap, min}$, implying that $t_{\rm 24\mu m}$ can be constrained with sufficient accuracy, but only upper limits can be derived for $\lambda$ and $t_{\rm fb,\,24\,\mu m}$.
\item The number of identified emission peaks is always above 35, both in the CO and the 24$\rm\,\mu m$ emission maps, as well as in the H$\alpha$ maps of IC\,342 and M31. 
\item The measured gas-to-SFR flux ratios focusing on gas (SFR tracer) peaks should never be below (above) the galactic average. This condition may not always be true in the presence of a diffuse emission reservoir. As visible in Figure~\ref{fig:tuningfork}, this criterion is satisfied after we filter out the large-scale diffuse emission in both tracer maps.
\item In order to perform accurate measurements, we require the global star formation history to not vary more than $0.2$~dex during the duration of the whole evolutionary cycle (ranging from $\sim$15 to 35~Myr for our sample) when averaged over time intervals of width $t_{\rm 24\,\mu m}$ or $t_{\rm CO}$. Using multi-wavelength data of the LMC and synthetic stellar population models, \citet{harris09} studied the star formation history in the LMC and found a roughly consistent SFR during the duration of the whole evolutionary cycle. The criterion is also satisfied in M33  and NGC\,300 as shown by \citet{kang12, kang16} using chemical evolution models to reconstruct the star formation rate history. Using data from the PHAT survey, the recent star formation history of M31 is confirmed to be quiescent without significant variations \citep{lewis15, williams17}. By performing spectral energy distribution fitting to the multi-wavelength data of M51, the SFR in M51 is also measured to be roughly constant for the last 100~Myr \citep{eufrasio17}. The star formation history of IC\,342 is not known. However, we do not expect it to experience significant variations in the global star formation rate during the last $\sim$30~Myr when averaged over $t_{\rm 24\,\mu m}=8$~Myr. In addition, we mask the starburst nucleus of this galaxy, which experienced a major burst of star formation $\sim$60~Myr ago \citep{boker99}.
\item Each independent region should be detectable at given sensitivity in both tracers at some point in their life. In order to check if this condition is satisfied in our sample of galaxies, we first calculate the minimum star-forming region mass expected to form from the detected molecular clouds by multiplying the star formation efficiency obtained in our method (see Table~\ref{tab:results}) by the 5$\sigma$ point-source sensitivity limit of the CO map. We then compare this mass to the mass of the stellar population required to provide an ionizing radiation luminosity that matches the 5$\sigma$ sensitivities of 24$\rm\,\mu m$ and H$\alpha$ maps on the scale of individual star-forming regions ($\lambda$). We use the Starburst99 model \citep{leitherer99} to calculate the initial mass of the stellar population assuming stars formed instantaneously 5~Myr ago (similarly to the H$\alpha$ emitting timescale). Since Starburst99 only provides models for the H$\alpha$ luminosity as a function of the age of the stellar population at different metallicities, we use the relation from \citet[][and references therein]{kennicutt12}, $\log\,{\rm SFR} = \log\,(L_{\rm H\alpha}) - 41.27 = \log\,(\rm\nu \textit{L}_{24\,\mu m}) - 42.69$, to obtain a similar estimation of the 24$\rm\,\mu m$ luminosity. We find that the minimum mass of the stellar population obtained from CO maps agrees well with that obtained from 24$\rm\,\mu m$ and H$\alpha$ maps (ranging from $100~\rm M_{\odot}$ to $5000~\rm M_{\odot}$ for the galaxies in our sample), suggesting that the sensitivity of the gas and SFR tracer maps are well-matched and the faintest CO peak is likely to evolve into the faintest \HII region. In principle, clouds can disperse dynamically before forming massive stars and then reassemble. In this case, the time spent before GMC dispersal would be added by our method to the measured lifetimes of clouds that do form massive stars. However, \citet{kruijssen19} and \citet{chevance20} show that this is unlikely to happen because the clouds are found to live only for about one dynamical timescale, not leaving enough time for clouds to disperse and recollapse before forming massive stars.
\end{enumerate}

Most of our measurements of $t_{\rm 24 \,\mu m}$ and $\lambda$ for all the galaxies in our sample as well as the $t_{\rm CO}$ and $\lambda$ for IC\,342 and M31 with H$\alpha$ as an SFR tracer are validated by satisfying the conditions listed above. The only exception is for M51, where we neither have sufficient resolution to accurately constrain the region separation length nor the feedback timescale. Only upper limits can be obtained for these values. In order to determine whether our measurements are reliable for $t_{\rm fb,\,24\,\mu m}$ for all the galaxies in our sample and $t_{\rm fb, H\alpha}$ for IC\,342 and M31, we use four additional criteria listed in \citet{kruijssen18}. To do so, we first introduce the filling factor of SFR tracer or gas peaks as $\zeta = 2r/\lambda$, where $r$ is the mean radius of the corresponding peaks. This parameter characterises how densely the peaks are distributed in a map. 

\begin{figure}
\includegraphics[width=\linewidth]{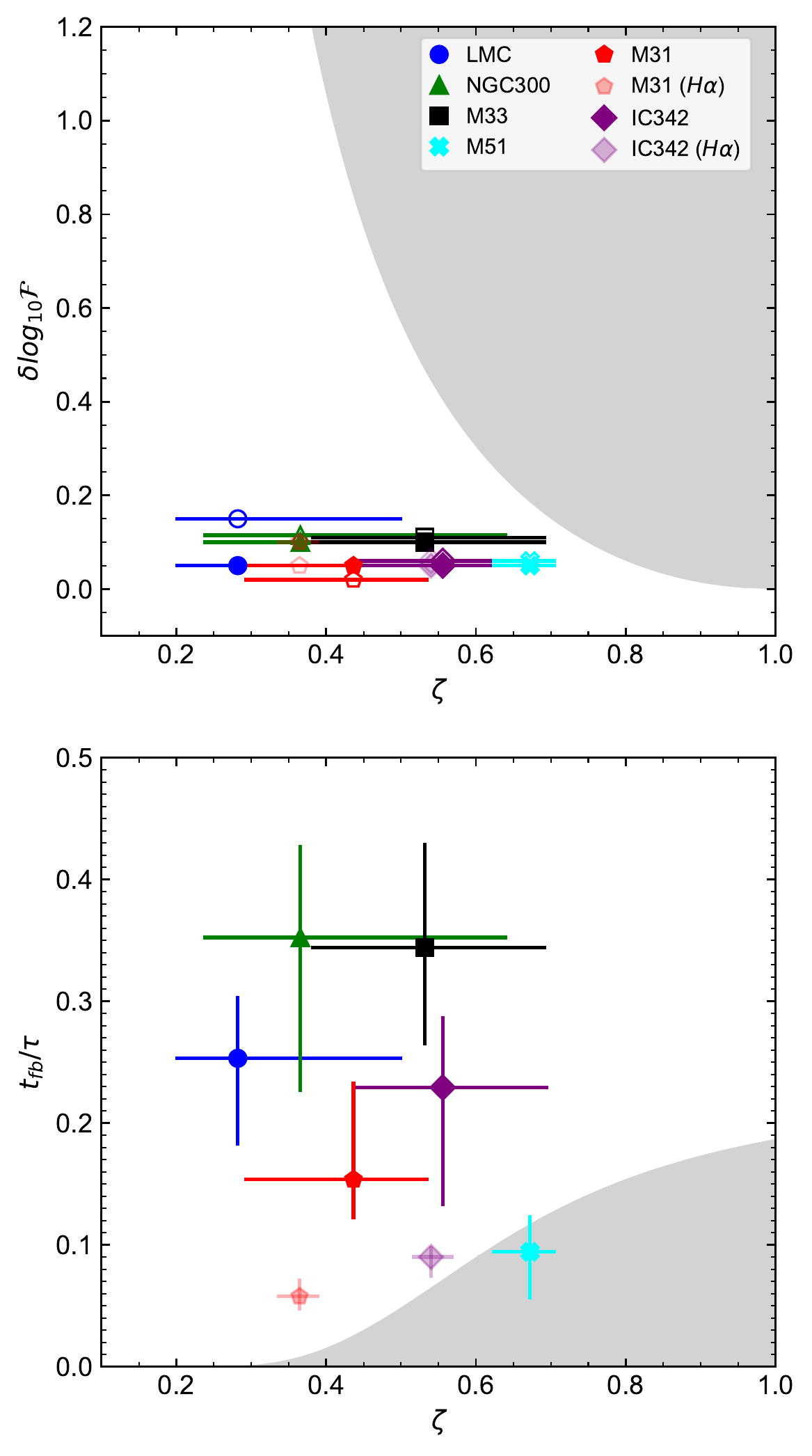}
\caption{Effects of blending on the feedback timescale measurements. The top panel shows the adopted density contrasts ($\delta\rm log_{10}\mathcal{F}$) used for peak identification in each 24$\rm\,\mu m$ (filled symbols) and CO (open symbols) emission map, as a function of the average filling factor $\zeta$. The transparent markers for IC\,342 and M31 indicate the adopted $\delta\rm log_{10}\mathcal{F}$ and measured $\zeta$ for the analysis with H$\alpha$ as SFR tracer. The shaded area indicates the region of the parameter space where peak identification is affected by blending \citep{kruijssen18}. Our results are well outside of the shaded area, confirming that we adopt small enough $\delta\rm log_{10}\mathcal{F}$ to identify adjacent peaks even in maps with high filling factor. The bottom panel shows the ratio between the feedback timescale ($t_{\rm fb}$) and the total duration of the whole evolutionary cycle ($\tau$) as a function of the average filling factor. The grey shaded area indicates the region of the parameter space where the contamination by neighbouring peaks affects the measurement of the feedback time. As a result, only an upper limit can be determined for $t_{\rm fb,\,24\,\mu m}$ in M51, whereas accurate measurements can be made for all other galaxies.}
\label{fig:blending}
\end{figure}

\begin{enumerate}\addtocounter{enumi}{6}
\item If peaks are densely distributed and potentially overlapping with each other, the density contrast used for peak identification ($\delta\rm log_{10}\mathcal{F}$) should be small enough to identify adjacent peaks. We compare in Figure~\ref{fig:blending} our values for $\delta\rm log_{10}\mathcal{F}$ with the upper limit prescribed by \citet{kruijssen18} and show that our choice enables the appropriate detection of neighbouring peaks, even in densely populated environments. 
\item For an accurate measurement of the feedback timescale, contamination by neighboring peaks should be small enough. Indeed, spatial overlap of neighbouring peaks due to low resolution or blending can be falsely attributed to a temporal overlap, therefore artificially increasing the duration of the measured feedback timescale. In this case, only an upper limit on the feedback timescale can be determined. In Figure~\ref{fig:blending}, we compare the analytical prescription of \citet{kruijssen18} with our measurements of $t_{\rm fb}/\rm \tau$ and average $\zeta$, where $\tau$ is the total duration of the whole evolutionary cycle ($\tau=t_{\rm CO}+t_{\rm 24\,\mu m}-t_{\rm fb,\,24\,\mu m}$). The average $\zeta$ is obtained by weighting the filling factors for gas and SFR tracer peaks by their corresponding timescales. We find that this condition is not satisfied for $t_{\rm fb,\,24\,\mu m}$ in M51. Only upper limits on this quantity can be determined (see Table~\ref{tab:results}). 
\item Figure~\ref{fig:blending} shows that the conditions $t_{\rm fb} > 0.05 \tau$ and $t_{\rm fb} < 0.95 \tau$ are verified for all galaxies.
\item Similarly to condition (v), the SFR should not vary more than $0.2$~dex during the entire timeline when averaged over the width of feedback timescale. This condition is also satisfied using the same reasoning, as stated in (v) above. 
\item After masking obvious blended regions such as galactic centres, visual inspection of the maps does not reveal abundant blending (Figure~\ref{fig:map_rgb}).
\end{enumerate}

Overall, we find that our measurements are reliable except for $\lambda$ and $t_{\rm fb,\,24\,\mu m}$ in M51. These two measurements should formally be considered as upper limits as they do not satisfy conditions (ii) and (viii). However, we note that  the deconvolution artifacts present in the 24$\rm\,\mu m$ map of M51 may (or may not) bias the feedback timescale and therefore the value we obtain as the upper limit should be considered uncertain.

\subsubsection{Effect of spatial resolution and inclination on the measured quantities}\label{subsec:robust_dist}

In order to test the potential effect of spatial resolution on our measurements, we have degraded the resolution of the CO and 24$\rm\,\mu m$ emission maps of NGC\,300 to the coarsest resolution among our galaxy sample (107~pc, see Table~\ref{tab:obs}) and repeated the analysis described in Section~\ref{sec:method}. We choose NGC\,300 for this test as it does not require bright star-forming regions to be masked like in the LMC and M33, making the application of the method more straightforward. We find that timescales ($t_{24\,\mu m}=8.4^{+3.8}_{-2.1}$~Myr; $t_{\rm fb,\,24\,\mu m}=4.8^{+1.5}_{-2.6}$~Myr) and the mean separation length ($\lambda=217^{+145}_{-112}$~pc) measured at a different spatial resolution are consistent within 1$\sigma$ uncertainties with the results of Table~\ref{tab:results}, confirming previous resolution tests on simulated galaxies \citep{kruijssen18} and on NGC\,300 \citep{kruijssen19} using H$\alpha$ as a SFR tracer.

We have also examined the possible effect of inclination on our measurements by repeating the analysis on CO and 24$\rm\,\mu m$ emission maps of NGC\,300, which have been artificially inclined further to match the highest inclination angle among our galaxy sample (M31; $i=77.7^{\circ}$). We find that measured timescales ($t_{24\,\mu m}=10.4^{+2.7}_{-2.0}$~Myr; $t_{\rm fb,\,24\,\mu m}=3.9^{+1.0}_{-0.9}$~Myr) and the mean separation length ($\lambda=192^{+130}_{-60}$~pc) agree within 1$\sigma$ uncertainties with those of NGC\,300 from Table~\ref{tab:results}. This is already expected from a similar test performed by \citet{kruijssen18} using simulated galaxies, where our method has been shown to provide reliable measurements even for a highly inclined galaxy as long as independent star-forming regions are sufficiently resolved ($\lambda \geq 1.5 l_{\rm ap, min}$; see (ii) in Section~\ref{subsec:robust_kl18}).

\subsection{Comparison with other works}\label{subsec:comparison}
The duration of the embedded star-forming phase has been measured in M33 by \citet{corbelli17}, using IRAM CO data and the mid-infrared source catalog created by \citet{sharma11}. In their work, GMCs and star-forming regions are classified into different evolutionary stages based on the presence of CO emission and SFR tracers such as 24$\rm\,\mu m$ and H$\alpha$ or UV emission. The clouds are defined to be in an inactive stage when no sign of star formation is detected, an embedded star-forming phase when CO emission is observed in association with 24$\rm\,\mu m$ but without associated H$\alpha$ or FUV emission. The region is defined to be at an exposed star-forming phase when H$\alpha$ or FUV emission becomes visible. 

The age estimates of the exposed star-forming regions (referred to as C-type in \citealp{corbelli17}) from spectral energy distribution (SED) fitting are available in \citet{sharma11}, and are obtained using photometric data at various wavelengths simultaneously, such as UV, H$\alpha$, and 24$\rm\,\mu m$. The age of the C-type phase corresponds to the time it takes for the cloud to evolve from the end of the heavily obscured phase of star formation (observed with 24$\rm\,\mu m$ but without H$\alpha$) to the end of the exposed young stellar region phase (both 24$\rm\,\mu m$ and H$\alpha$ are observed). This duration therefore corresponds to $ t_{\rm 24\,\mu m}-t_{\rm obscured}$ in our analysis. \citet{corbelli17} find that the C-type phase in M33 lasts for 8~Myr (without quoted uncertainty), which is in excellent agreement with our measurement of $8.4_{-3.0}^{+3.6}$~Myr. For the duration of the heavily obscured phase of star formation (CO and 24$\rm\,\mu m$ emission without H$\alpha$; referred to as B-type), \citet{corbelli17} find $2.4$~Myr, which is similar to the duration we measure, not only for M33 ($3.5_{-1.9}^{+1.2}$~Myr), but for most of the galaxies in our sample ($1.4-3.8$~Myr; see Section~\ref{subsec:em}). Lastly, as for the duration of the inactive phase (referred to as A-type), we measure $t_{\rm CO}- t_{\rm fb,\,24\,\mu m}=7.7_{-1.7}^{+1.4}~\rm Myr$, which is somewhat longer than the measurement of 4~Myr from \citet{corbelli17}. However, given the uncertainties in age estimates using SED fitting (on the order of $0.1$~dex) and the absence of any uncertainties on their estimates, the evolutionary timeline of molecular clouds of M33 from \citet{corbelli17} and our analysis are in good agreement.

The time it takes for the star-forming regions to become exposed has also been measured using wavelengths other than 24$\rm\,\mu m$ as a tracer for the embedded star formation. \citet{calzetti15} have measured ages of young massive star clusters in the dwarf starburst galaxy NGC\,5253 by applying SED modeling techniques on UV-optical-near infrared \textit{Hubble Space Telescope} photometry. While the star clusters have ages spanning from 1 to 15~Myr, the age estimate of one very heavily attenuated cluster with a clear near-infrared excess indicates that the duration of the heavily obscured phase of star formation is longer than (or similar to) 1~Myr for this particular star-forming region. \citet{whitmore14} used free-free radio continuum emission to detect heavily obscured star-forming regions, and charaterised the evolutionary timeline from quiescent molecular clouds to exposed star-forming phase using age estimates from SED fitting of young stellar regions in the overlap region of the merging Antennae galaxies. The duration of the heavily obscured phase (referred to as Stage~2 in \citealt{whitmore14}) and the feedback timescale (including the embedded phase; referred to as Stage~3 in \citealt{whitmore14}) are measured to be $0.1-1$~Myr and $1-3$~Myr, somewhat shorter than the duration we measure with 24$\rm\,\mu m$, which are $1-4$~Myr and $2-7$~Myr, respectively. We note that this difference could be because (i) the measurements are for galaxies undergoing a merger, unlike our sample; (ii) a different tracer is used to trace embedded star formation; and (iii) age estimates in highly extincted regions have considerable uncertainties \citep{hollyhead15}.

In conclusion, despite differences in methods, wavelengths, and galaxies used when constraining the evolutionary cycle of star-forming regions, our results are in good agreement with the measured timescales for the heavily obscured phase and feedback phase found in previous literature. The key step made in the present paper is to generalise these results to a sample of five galaxies (except M51), analysed homogeneously with a single analysis framework that is agnostic about which entities constitute a GMC or star-forming region.

\subsection{Effects of infrared emission not associated with local recent massive star formation}\label{subsec:eff_evolved}

24$\rm\,\mu m$ emission is widely used as a tracer for embedded star formation, as it captures emission of massive stars that has been reprocessed by dust grains \citep[see e.g.][]{calzetti07, kennicutt12, vutisalchavakul13}. However, one of the known issues with using 24$\rm\,\mu m$ emission to trace recent star formation is that the interstellar radiation field, late-type B~stars (age of $\sim$100~Myr), and dust clumps heated by external radiation such as nearby star-forming regions also contribute to the emission at this wavelength \citep{ calzetti07, murphy11,  kennicutt12, leroy12}. 

The difference in spatial distributions associated with each process generating 24$\rm\,\mu m$ emission allows us to separate the emission for recent star formation events from other sources. The 24$\rm\,\mu m$ emission originating from the interstellar radiation field has an extended morphology, because it originates from small dust grains in the diffuse interstellar medium \citep{draine07, draine07b, verley09, rahman11, leroy12}. Such diffuse emission, constituting on average of 55\% of the 24$\rm\,\mu m$ emission, is therefore expected to be removed during our filtering process. 

On the other hand, the effect of 24$\rm\,\mu m$ emission associated with late-type B~stars and starless dust clumps might not be filtered out because they are more inhomogeneously distributed, similarly to the emission from young star-forming regions. To estimate the effect of late-type B~stars and externally illuminated dust clumps on our measurements, we make use of the far-infrared source catalogue of the LMC provided by \citet{seale14}. In this catalogue, young stellar objects and dust clumps (that may or may not have deeply embedded forming stars) are identified, as well as sources not related to recent star formation such as asymptotic giant branch stars, planetary nebulae, and supernova remnants using literature catalogs (\citealt{seale14} and references therein). In order to test whether the inclusion of 24$\rm\,\mu m$ emission from older stars and dust clumps could bias our results, we mask these sources and repeat our analysis of the LMC. When masking dust clumps, we mask all the probable candidates in \citet{seale14} as it is difficult to distinguish whether these clumps harbour deeply embedded stars or are heated by external radiation. We find that the older stars and dust clumps have a negligible effect on our results. In practice, older stars and dust clumps are not usually identified as SFR tracer peaks in our analysis due to their low brightness and small size, which does not satisfy the requirement of a minimum number of pixels to be identified as a peak in our method.

In conclusion, once the diffuse emission has been filtered, the 24$\rm\,\mu m$ maps mostly contain emission from young stars. Any potential bias due to the interstellar radiation field, late-type B~stars, and dust clumps is negligible and our measurements of $t_{\rm24\,\mu m}$ provide an accurate characterisation of the duration of (partially) embedded massive stars.

\section{Conclusion}\label{sec:conclusion}
We present a characterisation of the evolutionary timeline from molecular clouds to young stellar regions in six nearby galaxies by applying the statistical method developed by \citet{kruijssen14} and \citet{kruijssen18} to CO and 24$\rm\,\mu m$ emission maps at cloud-scale ($20{-}100$~pc) resolution. With this method, we measure the duration of the 24$\rm\,\mu m$ emission phase ($t_{24\,\mu m}$), the duration of the feedback phase ($t_{\rm fb,\,24\,\mu m}$) during which massive star formation continues embedded in molecular clouds, the duration of the heavily obscured star formation phase with no associated H$\alpha$ emission ($t_{\rm obscured}$), and the average distance between independent star-forming regions evolving from clouds to massive star formation ($\lambda$). We also derive other physical quantities such as the feedback velocity ($v_{\rm fb}$) and the integrated star formation efficiency per star formation event ($\epsilon_{\rm sf}$) from our measurements.

Across our sample of galaxies, we find that molecular clouds are quickly disrupted within $2-7$~Myr after the onset of embedded massive star formation (traced by 24~$\mu$m emission) by stellar feedback, supporting the fact that GMCs are dispersed within a cloud dynamical timescale, as suggested by \citet{elmegreen00} and \citet{hartmann01}. The measured feedback timescale, which includes the duration of the massive star-forming phase, constitutes $17-47$\% of the cloud lifetime of $10-30$~Myr. The feedback timescales are generally shorter than the time it takes for the first supernova to explode ($4-20$~Myr), when stochasticity of the initial mass function is taken into account \citep{chevance20_fb}, suggesting that early feedback mechanisms such as photoionization and stellar winds are mainly responsible for the dispersal of molecular clouds. Previous works have found similar duration of this phase using age estimates of star clusters in the Milky Way and some nearby galaxies \citep{lada03, whitmore14, corbelli17}. After the molecular gas is dispersed, the 24$\rm\,\mu m$ emission decays within $2-9$~Myr. Our results further support the conclusion of earlier work that galaxies are composed of independent star-forming regions separated by ${\sim}{100-200}$~pc \citep{kruijssen19, chevance20}, which may correspond to the vertical gas disc scale height (see \citealp{kruijssen19}). These regions are undergoing an inefficient star-forming process with integrated cloud-scale star formation efficiencies ($\epsilon_{\rm sf}$) of $0.7-6.8$\%. The measured star formation efficiencies are consistent with previous measurements in these galaxies using other tracers to estimate the global SFR. We obtain feedback velocities ($v_{\rm fb}$) of $8-30$ \kms, which is consistent with the observed expansion velocities of nearby \HII regions \citep[e.g.][]{murray10,mcleod19,mcleod20,barnes20}.

By combining our measurements with those using H$\alpha$ as a tracer for exposed star-forming regions, we also measure the duration of the heavily obscured phase (detected with CO and 24$\rm\,\mu m$ but without H$\alpha$ emission). Our results show that this period lasts for $3.0\pm0.9$~Myr (with a full range of $1.4-3.8$~Myr across our sample of galaxies). We do not detect any significant correlation of the duration of this heavily obscured phase with galactic properties. This measured duration is in good agreement with values suggested by previous works using different wavelengths, methods, and galaxies.

Furthermore, we study the correlation of our measurements with galactic (or cloud-scale) properties, such as mass-weighted mean surface density of GMCs, metallicity, star formation efficiency, SFR surface density, and the feedback velocity. While we do not find statistically significant trends, the durations of the total and isolated 24$\rm\,\mu m$ emission phases ($t_{\rm 24\,\mu m}$ and  $t_{\rm iso, 24\,\mu m}$) may weakly decrease with increasing metallicity. We conjecture that this dependence results from winds of massive stars being stronger and more energetic at higher metallicities, which leads to a more effective dispersal of the clouds. No such trends with metallicity are observed for the feedback timescale and the duration of the heavily obscured phase.

In order to gain a better understanding of the mechanisms driving the early feedback process, a systematic measurement of the embedded phase in a large number of galaxies in various environments is essential. Due to the limited resolution of \textit{Spitzer} 24$\rm\,\mu m$ observations ($6.4\arcsec$), we have been able to perform this analysis for only six nearby galaxies and accurately constrain the duration of the embedded phase in five of them. In the future, the MIRI imager aboard the \textit{James Webb Space Telescope}, with a field of view of $1\arcmin \times 2 \arcmin$, will reach an angular resolution of $0.7\arcsec$. This will enable the application of the same method to galaxies located out to 25~Mpc, covering a much wider range of galaxy properties and morphologies, allowing us to explore how the feedback processes govern the evolution of molecular clouds during the early stages of star formation, as a function of the galactic environment.

\section*{Acknowledgements}
We thank an anonymous referee for helpful comments that improved the quality of the manuscript. We thank Alexander~Hygate for thankful suggestions and K.~Herrmann for kindly sharing the H$\alpha$ map of IC\,342. JK, MC, and JMDK gratefully acknowledge funding from the German Research Foundation (DFG) through the DFG Sachbeihilfe (grant number KR4801/2-1). MC and JMDK gratefully acknowledge funding from the DFG through an Emmy Noether Grant (grant number KR4801/1-1). JMDK gratefully acknowledges funding from the European Research Council (ERC) under the European Union's Horizon 2020 research and innovation programme via the ERC Starting Grant MUSTANG (grant agreement number 714907). JMDK gratefully acknowledge funding from from Sonderforschungsbereich SFB 881 (Project-ID 138713538) `The Milky Way System' (subproject B2) of the DFG. FB and ATB would like to acknowledge funding from the European Research Council (ERC) under the European Union’s Horizon 2020 research and innovation programme (grant agreement No.726384/Empire). SCOG and RSK acknowledges support from the Deutsche Forschungsgemeinschaft (DFG) via the Collaborative Research Center (SFB 881, Project-ID 138713538) ``The Milky Way System'' (sub-projects A1, B1, B2 and B8) and from the Heidelberg cluster of excellence (EXC 2181 - 390900948) ``STRUCTURES: A unifying approach to emergent phenomena in the physical world, mathematics, and complex dat'', funded by the German Excellence Strategy. They also thank for funding form the European Research Council in the ERC Synergy Grant ``ECOGAL -- Understanding our Galactic ecosystem: From the disk of the Milky Way to the formation sites of stars and planets'' (project ID 855130). KK gratefully acknowledges funding from the German Research Foundation (DFG) in the form of an Emmy Noether Research Group (grant number KR4598/2-1, PI Kreckel). The work of AKL and JS is partially supported by the National Science Foundation (NSF) under grant Nos.\ 1615105, 1615109, and 1653300. MQ acknowledges support from the research project PID2019-106027GA-C44 from the Spanish Ministerio de Ciencia e Innovaci\'on. ES and TGW acknowledge funding from the European Research Council (ERC) under the European Union’s Horizon 2020 research and innovation programme (grant agreement No. 694343). This work was carried out as part of the PHANGS collaboration. This paper makes use of the following ALMA data: ADS/JAO.ALMA $\#$2013.1.00351.S, ADS/JAO.ALMA $\#$2015.1.00258.S. ALMA is a partnership of ESO (representing its member states), NSF (USA) and NINS (Japan), together with NRC (Canada), NSC and ASIAA (Taiwan), and KASI (Republic of Korea), in cooperation with the Republic of Chile. The Joint ALMA Observatory is operated by ESO, AUI/NRAO, and NAOJ. The National Radio Astronomy Observatory is a facility of the National Science Foundation operated under cooperative agreement by Associated Universities, Inc. This work makes use of the PdBI Arcsecond Whirlpool Survey (Pety et al. 2013; Schinnerer et al. 2013). The authors thank IRAM for making the data products of IC\,342 \citep{schruba21_ic342}, M31 \citep{schruba21_m31}, and M33 CO Large Program \citep{gratier10, druard14} available. IRAM is supported by INSU/CNRS (France), MPG (Germany), and IGN (Spain). We thank the IRAM staff for their assistance with the observations.
\vspace{-4mm}

\section*{Data availability}
The data underlying this article will be shared on reasonable request to the corresponding author.

\bibliographystyle{mnras}
\bibliography{mybib}
\appendix

\section{Individual images of emission maps used in our analysis}\label{sec:app_map}
\begin{figure*}
\includegraphics[width=16cm,height=21.3cm]{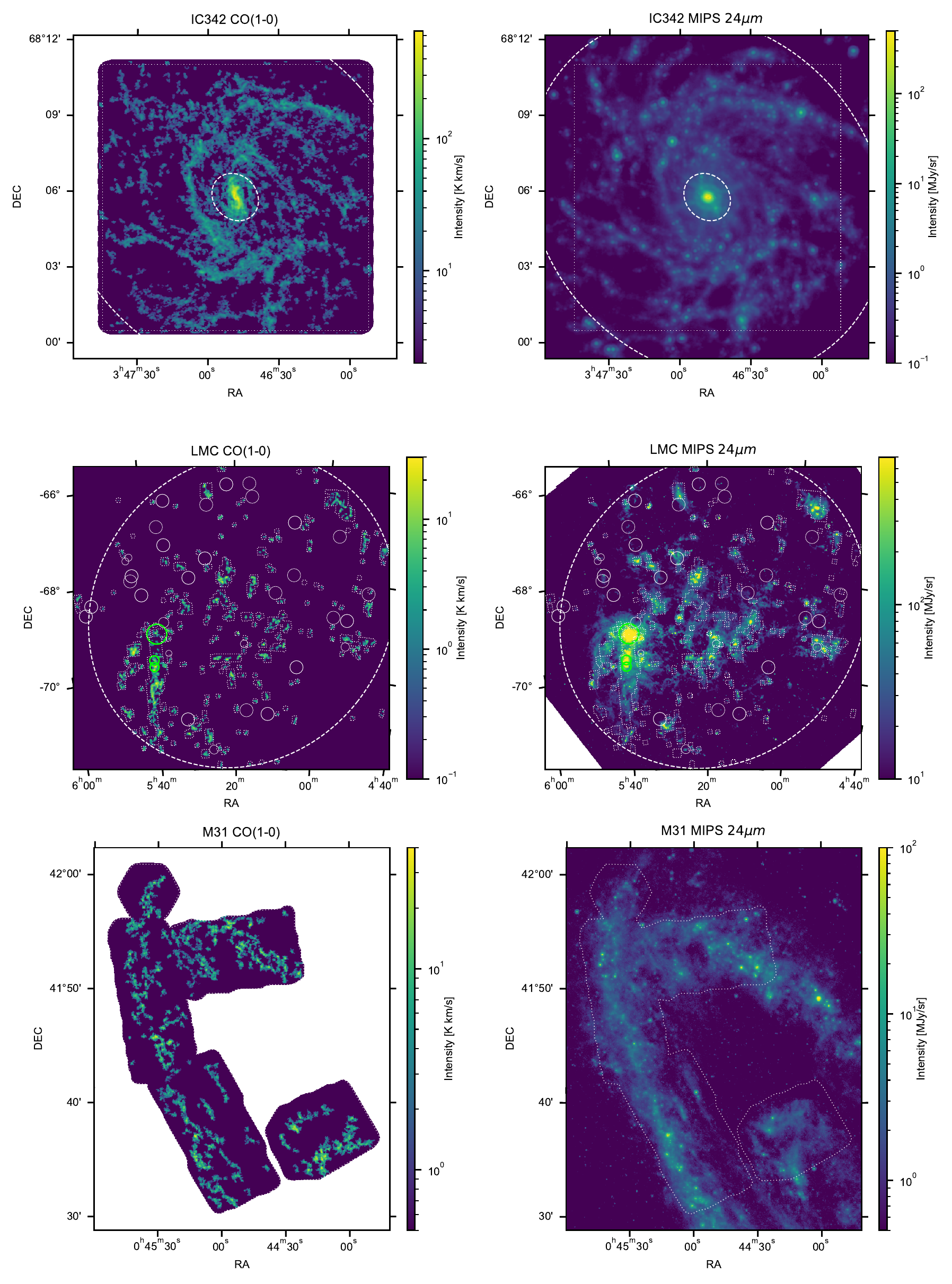}
\caption{Maps of \coone emission (left panels) and 24$\rm\,\mu m$ emission (right panels) for the IC\,342, LMC, and M31. The ranges of galactic radii included in our analysis are indicated by white dashed ellipses. The white dotted line shows the coverage of the CO observations. In the LMC, regions where molecular gas exists but was not targeted by the MAGMA survey are masked (white solid circles). The masked bright star-forming regions are shown by green circles. We also mask foreground stars, background galaxies, and map artifacts (purple circles)} \label{fig:app_map1}
\end{figure*}

\begin{figure*}
\includegraphics[width=\textwidth, height=22.2cm]{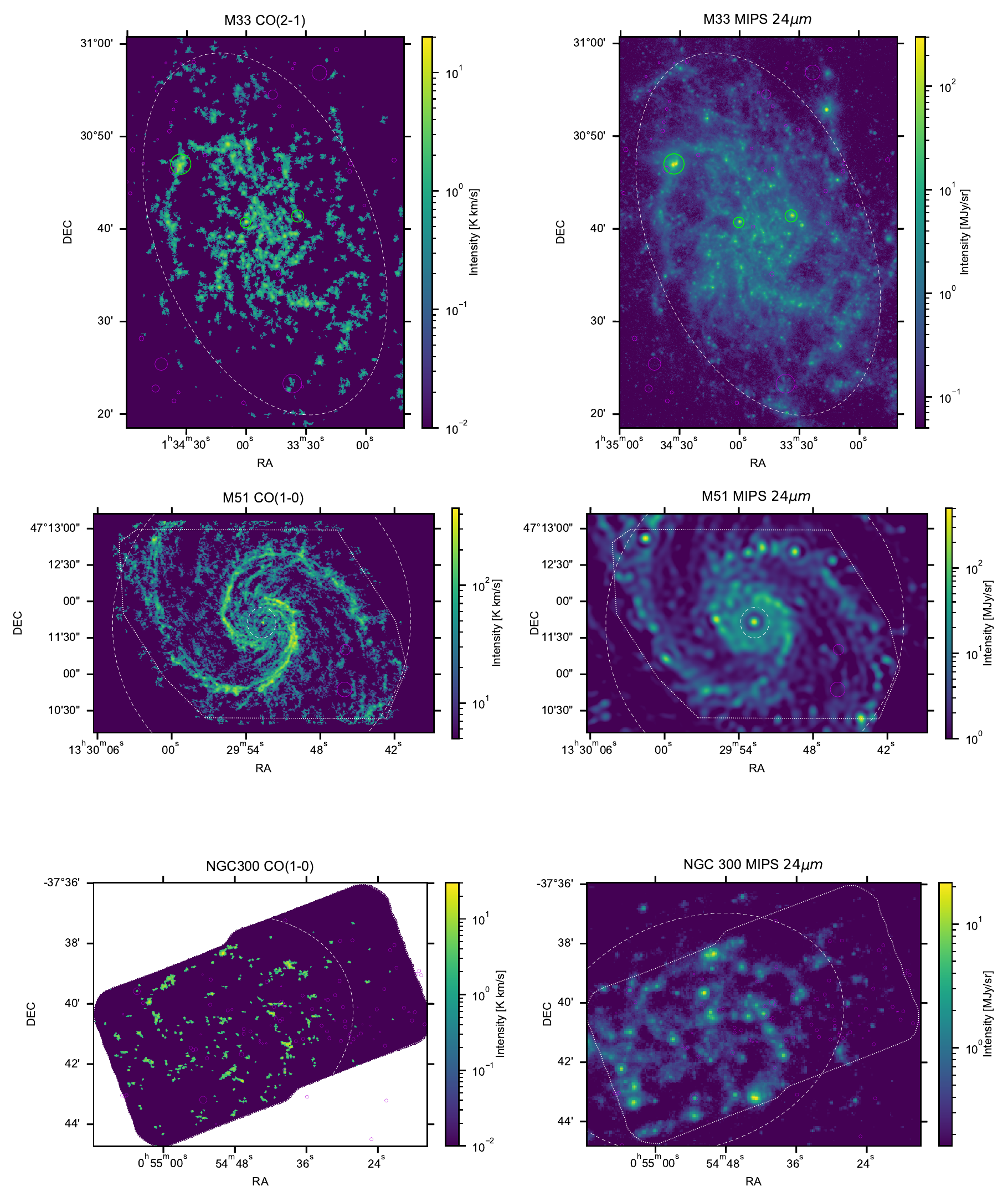}
\caption{Same as Figure~\ref{fig:app_map1}, but here for M33, M51, and NGC\,300. Integrated intensity maps of \coone are shown for M51 and NGC\,300, while \cotwo is shown for M33.}\label{fig:app_map2}
\end{figure*}

In Figures~\ref{fig:app_map1} and \ref{fig:app_map2}, we present the CO and 24$\rm\,\mu m$ emission maps used in our analysis to trace molecular gas and young stellar regions, respectively.  

\section{Molecular cloud lifetimes in IC\,342 and M31}\label{sec:app_tco}

\begin{figure}
\includegraphics[width=\linewidth]{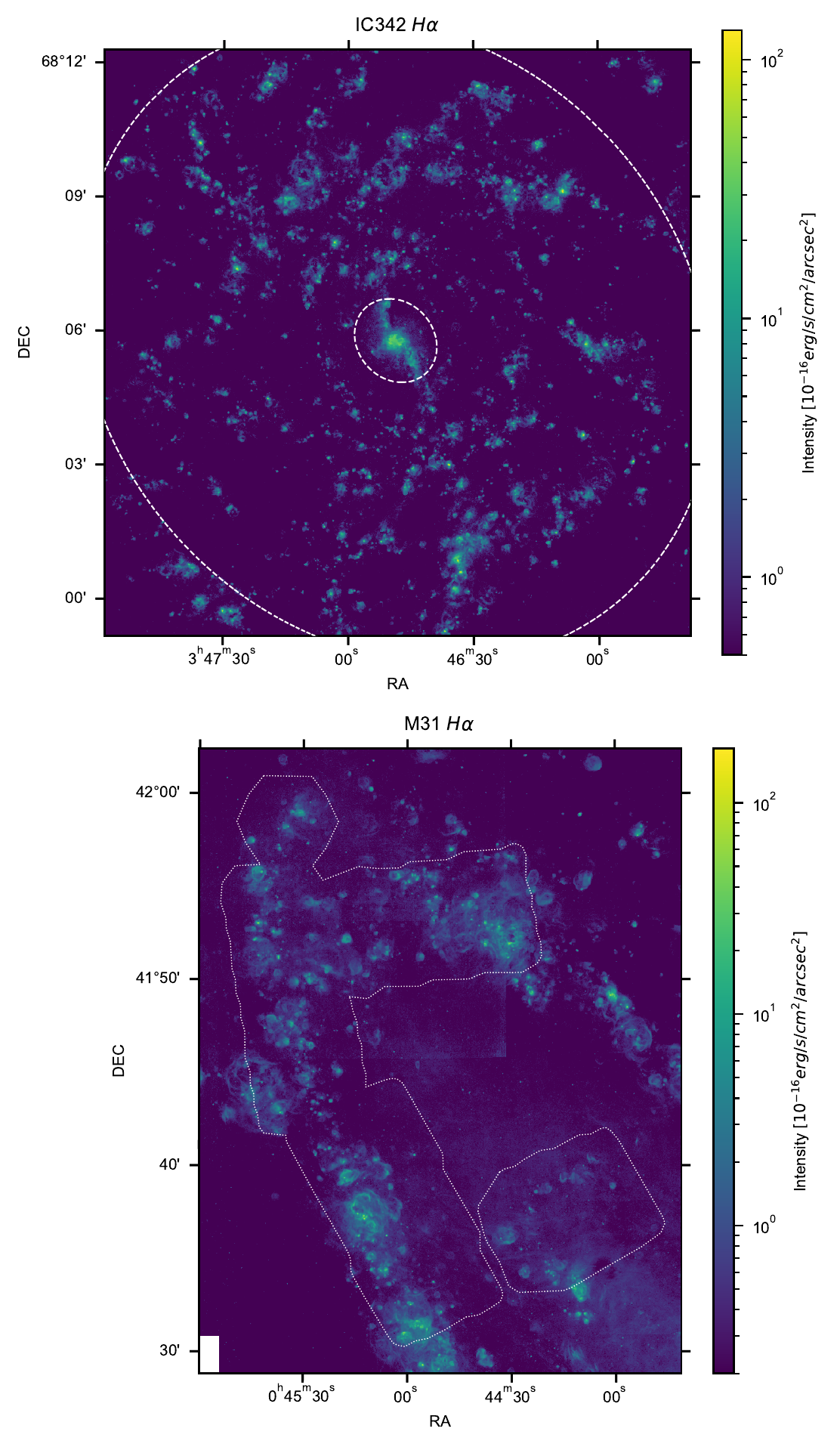}
\caption{Same as Figure~\ref{fig:app_map2}, for H$\alpha$ emission maps of IC\,342 and M31.}
\label{fig:app_ha}
\end{figure}

In order to derive absolute durations of the different phases of cloud evolution and star formation (see Section~\ref{sec:method}), we have used the cloud lifetime ($t_{\rm CO}$) as the
reference timescale in our analysis (see Table~\ref{tab:input}). For four out of six galaxies in our sample, the cloud lifetime has been constrained in previous works using H$\alpha$ as a tracer for exposed star-forming regions. Here, we describe the characterization of the cloud lifecycle of IC\,342 and M31, by applying the same method as described in Section~\ref{sec:method} using CO and H$\alpha$ emission maps to trace molecular gas and young stellar regions, respectively. 

Figure~\ref{fig:app_ha} shows the H$\alpha$ emission maps of IC\,342 and M31 used to trace young massive star-forming regions. A summary of the observational data is presented in Section~\ref{subsec:datasets}. Table~\ref{tab:app_input} lists the adopted main input parameters specific to the analysis using CO and H$\alpha$ as molecular gas and SFR tracers, respectively. Other input parameters are listed in Table~\ref{tab:obs} and Table~\ref{tab:input}. We set the minimum aperture size ($l_{\rm ap, min}$) to match the CO map resolution as it is coarser compared to the H$\alpha$ map resolution. The reference timescales ($t_{\rm ref}$) for the exposed young stellar phase (duration of the isolated H$\alpha$ emission phase) are adopted from \citet{haydon20}. The duration does not include the feedback timescale, so the total duration of the H$\alpha$ emission phase ($t_{\rm H\alpha}$) equals $t_{\rm ref}+t_{\rm fb,\,H\alpha}$.

\begin{figure}
\centering
\includegraphics[width=0.8\linewidth]{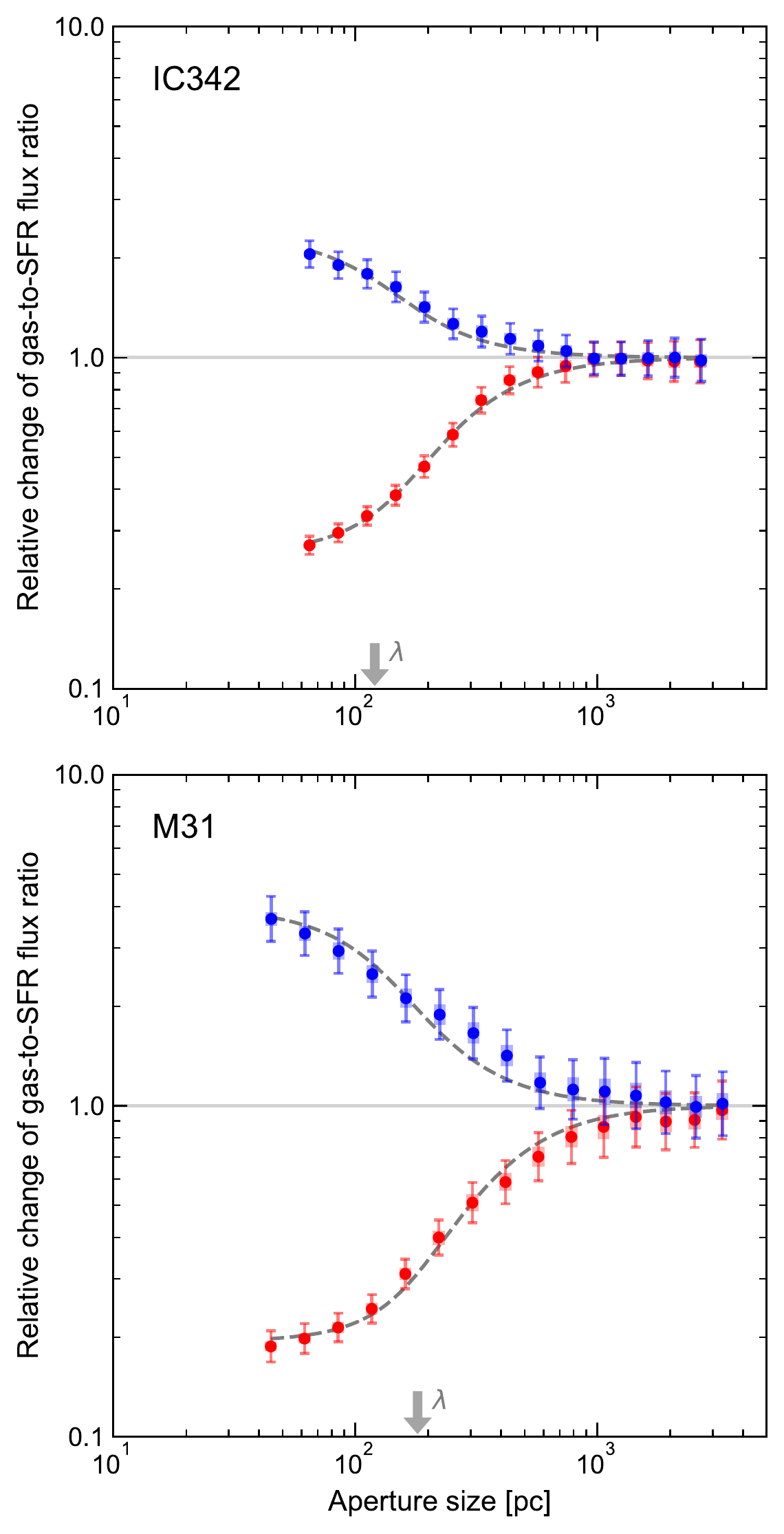}
\caption{Relative change of the gas-to-SFR (CO-to-H$\alpha$) flux ratio compared to the galactic average as a function of size of apertures placed on CO (blue) and H$\alpha$ (red) emission peaks. The error bars indicate 1$\sigma$ uncertainty on each individual data point whereas the shaded area is an effective 1$\sigma$ uncertainty taking into account the covariance between data points. Our best-fitting model is shown as dashed line and the solid horizontal line indicates the galactic average. The measured region separation length ($\lambda$) is indicated in each panel and other constrained best-fitting parameters ($t_{\rm CO}$ and $t_{\rm fb,\,H\alpha}$) are listed in Table~\ref{tab:results}.}
\label{fig:app_tuningfork_coha}
\end{figure}

In Figure~\ref{fig:app_tuningfork_coha}, we show the measured deviations of the enclosed gas-to-SFR flux ratios in apertures centred on CO and H$\alpha$ peaks, compared to the galactic average, together with our best-fitting model. Table~\ref{tab:results} lists the constrained quantities for the best-fitting model. We have verified the accuracy of these measurements in Section~\ref{subsec:robust}. As seen in our results using 24$\rm\,\mu m$ as an SFR tracer, we find a spatial de-correlation between gas and SFR tracer emission peaks. We measure cloud lifetimes of $20.0_{-2.3}^{+2.0}$~Myr for IC\,342 and $14.0_{-1.9}^{+2.1}$~Myr for M31. The measured cloud lifetimes are within the range of our previous measurements of other galaxies where GMCs are found to live for $10-30$~Myr \citep{kruijssen19, chevance20, hygatePhD, ward20, zabel20}. We suspect that the difference between the two cloud lifetimes are related to the different environments in which the molecular clouds are located, as recently suggested by \citet{chevance20}. In this case, the cloud evolution in IC\,342, which has a high molecular gas surface density ($\rm 9.55~M_{\odot}~pc^{-2}$), is likely to be governed by galactic dynamical processes, whereas internal dynamics such as free-fall and crossing times are the determinant factor for cloud evolution in low molecular gas surface density environments such as M31 ($\rm{\sim} 1~M_{\odot}~pc^{-2}$). The duration over which CO and H$\alpha$ emission overlap is short ($2.2_{-0.5}^{+0.4}$~Myr in IC\,342 and $1.1_{-0.2}^{+0.3}$~Myr in M31), as seen in previous measurements of other galaxies \citep{chevance20_fb}. These short feedback timescales indicate that molecular clouds are destroyed shortly after the star-forming region becomes exposed, making them visible in H$\alpha$. Finally, we find that independent star-forming regions are separated by $120_{-10}^{+10}$~pc in IC\,342 and $181_{-19}^{+28}$~pc in M31, comparable to our measurements with 24$\rm\,\mu m$ as an SFR tracer, as well as our previous findings with H$\alpha$ for different galaxies. 

\begin{table}
\begin{center}
\caption{Main input parameters of the analysis using H$\alpha$ as an SFR tracer for IC\,342 and M31. For other input parameters, we use the default values listed in table~2 of \citet{kruijssen18}. \label{tab:app_input}}
\begin{threeparttable}
\begin{tabular}{lcc}
\hline
Quantity & IC\,342 & M31 \\
\hline
$l_{\rm ap, min}$ [pc] 	&	 65 	&	45\\
$l_{\rm ap, max}$ [pc] 		&	 3000 	&	4000\\
$N_{\rm ap}$		&	 15  	&	15\\
$\rm N_{pix, min}$ 	&		 10  	&	20\\
$\Delta\rm log_{10}\mathcal{F}_{\rm CO}$ 	&		 1.1	&	1.2\\
$\delta\rm log_{10}\mathcal{F}_{\rm CO}$ 	&		 0.05 	&	0.05\\
$\Delta\rm log_{10}\mathcal{F}_{\rm H\alpha}$ 		&	 2.8 	&	2.0\\
$\delta\rm log_{10}\mathcal{F}_{\rm H\alpha}$ 		&	 0.05 	&	0.1\\
$t_{\rm ref}$ [Myr] 		&	 4.25 	&	4.42\\
$t_{\rm ref, errmin}$ [Myr] 	&		 0.15 	&	0.18\\
$t_{\rm ref, errmax}$ [Myr] 	&		 0.15 	&	0.19\\
$n_{\lambda}$&		 12 	&	10\\
\hline
\end{tabular}
\end{threeparttable}
\end{center}
\end{table}

\vspace{4mm}
\noindent {\it
$^1$Astronomisches Rechen-Institut, Zentrum f\"{u}r Astronomie der Universit\"{a}t Heidelberg, M\"{o}nchhofstra\ss e 12-14, 69120 Heidelberg, Germany\\
$^2$Max-Planck Institut f\"ur Extraterrestrische Physik, Giessenbachstra\ss e 1, 85748 Garching, Germany\\
$^3$Center for Astrophysics and Space Sciences, Department of Physics, University of California, San Diego, 9500 Gilman Drive, La Jolla, CA 92093, USA \\
$^{4}$Argelander-Institut f\"{u}r Astronomie, Universit\"{a}t Bonn, Auf dem H\"{u}gel 71, 53121 Bonn, Germany\\
$^{5}$The Observatories of the Carnegie Institution for Science, 813 Santa Barbara Street, Pasadena, CA 91101, USA\\
$^{6}$Departamento de Astronomía, Universidad de Chile, Casilla 36-D, Santiago, Chile\\
$^7$Aix Marseille Univ, CNRS, CNES, LAM (Laboratoire d’Astrophysique de Marseille), Marseille, France\\
$^8$Department of Physics \& Astronomy, University of Wyoming, Laramie, WY 8207\\
$^{9}$Department of Astronomy, University of Massachusetts - Amherst, 710 N. Pleasant St., Amherst, MA 01003\\
$^{10}$Instit\"ut  f\"{u}r Theoretische Astrophysik, Zentrum f\"{u}r Astronomie der Universit\"{a}t Heidelberg, Albert-Ueberle-Strasse 2, 69120 Heidelberg, Germany\\
$^{11}$Research School of Astronomy and Astrophysics, Australian National University, Canberra, ACT 2611, Australia\\
$^{12}$International Centre for Radio Astronomy Research, University of Western Australia, 7 Fairway, Crawley, 6009, WA, Australia\\
$^{13}$IRAM, 300 rue de la Piscine, 38406 Saint Martin d'H\`eres, France\\
$^{14}$Universit\"{a}t Heidelberg, Interdisziplin\"{a}res Zentrum f\"{u}r Wissenschaftliches Rechnen, Im Neuenheimer Feld 205, 69120 Heidelberg, Germany\\
$^{15}$Caltech/IPAC MC 314-6 (Keith Spalding Building)
1200 E California Blvd Pasadena, CA 91125\\
$^{16}$Department of Astronomy, The Ohio State University, 140 West 18th Ave, Columbus, OH 43210, USA\\
$^{17}$Sorbonne Universit\'e, Observatoire de Paris, Universit\'e PSL, CNRS, LERMA, F-75005, Paris, France\\
$^{18}$Observatorio Astron{\'o}mico Nacional (IGN), C/Alfonso XII 3, Madrid E-28014, Spain\\
$^{19}$Max Planck Institut für Astronomie, Königstuhl 17, 69117 Heidelberg, Germany\\
}

\bsp
\label{lastpage}
\end{document}